\documentclass[%
 amsmath,amssymb,
 aps, physrev,
]{revtex4-2}

\usepackage{graphicx}
\usepackage{dcolumn}
\usepackage{bm}

\usepackage{epstopdf, epsfig}
\usepackage{gensymb}
\usepackage{xcolor}
\usepackage[super]{nth}
\usepackage{multirow}
\usepackage{amstext} 
\usepackage{array}   
\usepackage{appendix}
\usepackage{booktabs}
\usepackage{amsmath}

\begin{document}

\title{\textbf{Asymptotic limits of the attached eddy model derived from an adiabatic atmosphere} 
}%

\author{Yue Qin}%
 \email{Contact author: yueqin@bu.edu}
\affiliation{Department of Earth and Environment, Boston University, Boston, Massachusetts, USA
}

\author{Gabriel G. Katul}
 
\affiliation{
 Department of Civil and Environmental Engineering, Duke University, Durham, North Carolina, USA
}%

\author{Heping Liu}
\affiliation{Department of Civil and Environmental Engineering, Washington State University, Pullman, Washington, USA
}%

\author{Dan Li}
\affiliation{Department of Earth and Environment, Boston University, Boston, Massachusetts, USA
}

\date{\today}

\begin{abstract}
The attached-eddy model (AEM) predicts mean velocity and streamwise velocity variance profiles that follow a logarithmic shape in the overlap region of high Reynolds number wall-bounded turbulent flows.  Moreover, the AEM coefficients are presumed to attain asymptotically constant values at very high Reynolds numbers. Here, the logarithmic behaviour of the AEM predictions in the near-neutral atmospheric surface layer is examined using sonic anemometer measurements from a 62-m meteorological tower located in the Eastern Snake River Plain, Idaho, US. Utilizing an extensive 210-day dataset, the inertial sublayer (ISL) is first identified by analyzing the measured momentum flux and mean velocity profile. The logarithmic behaviour of the streamwise velocity variance and the associated `-1' scaling of the streamwise velocity energy spectra are then investigated. The findings indicate that the Townsend-Perry coefficient ($A_1$) is influenced by mild non-stationarity that manifests itself as a Reynolds number dependence. After excluding non-stationary runs and requiring a Reynolds number higher than $4 \times 10^7$, the inferred $A_1$ converges to values ranging between 1 and 1.25, consistent with laboratory experiments. Moreover, the independence of the normalized vertical velocity variance from the wall-normal distance in the ISL is further checked and the constant coefficient value agrees with reported laboratory experiments at very high Reynolds numbers as well as many surface layer experiments.  Furthermore, nine benchmark cases selected through a restrictive quality control reveal a closer relationship between the `-1' scaling in the streamwise velocity energy spectrum and the logarithmic behaviour of streamwise velocity variance at higher Reynolds numbers, though no direct equivalence between the plateau value and $A_1$ is observed.

\end{abstract}
\maketitle

\section{Introduction}
The atmospheric surface layer (ASL), typically identified by the bottom 10\% of the atmospheric boundary layer (ABL), extends up to 50-100 metres above the ground.  It is a layer where the Coriolis effects are small and may be ignored in the mean momentum balance. Because the air kinematic viscosity $\nu$ is small but typical length and velocity scales associated with turbulence are large, the ASL can serve as a testing ground for assessing asymptotic convergence of laboratory experiments and theories in the limit where similarity coefficients become independent of Reynolds number. However, comparing the ASL to the much studied inertial sublayer (ISL) in canonical turbulent boundary layers of flumes, pipes, and wind tunnels is not free from challenges. Unlike laboratory settings where the boundary layer height is often known, the ABL height varies in time and space and is notoriously difficult to determine. Pragmatic issues such as achieving statistical convergence while ensuring stationarity arise due to the unsteadiness of ABL height. The ASL is also influenced by diurnal surface heating and cooling preventing a strict attainment of adiabatic conditions. Daytime conditions are characterized by higher winds but also experience higher surface heating complicating the full attainment of very large Reynolds numbers in a strict adiabatic manner. Moreover, flow statistics in the ASL are complicated by a plethora of other factors such as surface heterogeneity, terrain effects, and upstream influences making a zero-pressure gradient condition difficult to ensure \citep{Marusic2010}.  Despite these difficulties, several observational and comparative studies revealed that the ASL shares some similarities with the ISL of canonical wall-bounded incompressible flows such as zero-pressure gradient boundary layers and fully developed pipe and channel flow 
\citep{Smits2011,Hutchins2012, Marusic2013, Huang2022}. Long-term ASL experiments may offer a large ensemble of runs where a subset of those runs permits identifying conditions that match expectations from idealized laboratory studies.  

In the ISL of laboratory flows, one of the most cited models describing the velocity statistics at very high Reynolds numbers is the attached eddy model (AEM) originally put forth by Townsend \citep{Townsend1976}. For the mean streamwise velocity and streamwise velocity variance, AEM predicts 
\begin{equation}
\label{eqn:law_of_wall_1}
 \frac{\overline{u}}{u_*} = \frac{1}{\kappa} \ln \left(\frac{z-d}{z_0}\right),
\end{equation}
\begin{equation}
\label{eqn:law_of_wall_2}
 \frac{{\sigma}_u^2}{u_*^2} = B_1-A_1 \ln\left(\frac{z}{\delta}\right),
\end{equation}
where $u$ is the streamwise velocity component; $z$ is the wall-normal distance with $z=0$ set at the ground, $d$ is the zero-plane displacement, $z_0$ is the momentum roughness length; $\sigma_u^2=\overline{u'^2}$ is the variances of $u$, prime is fluctuation due to turbulence around the mean state; overline represents averaging over coordinates of statistical homogeneity (usually time); $u_*=(\tau/\rho)^{1/2}$ is the friction velocity assumed to be the normalizing velocity for the flow statistics in AEM with $\tau$ being the wall stress and $\rho$ is the density of the fluid and $\delta$ is the outer length scale of the flow approximating the boundary layer height. The coefficients $\kappa$ is the von K\'arm\'an constant, $A_1$ is often referred to as the Townsend-Perry coefficient, and $B_1$ is an empirical coefficient dependent on the flow (i.e. pipe flow versus wind tunnels).  The $\kappa$, $A_1$ and $B_1$ are presumed to attain asymptotic constant values at very large Reynolds numbers traditionally defined as $Re_{\tau}=u_*\delta/\nu$ \citep{Marusic2013, Smits2011}.
For the vertical velocity ($w$) variance ($\sigma_w^2=\overline{w'^2}$), AEM predicts
\begin{equation}
\label{eqn:law_of_wall_3}
 \frac{\sigma_w^2}{u_*^2}=B_2,
\end{equation}
where $B_2$ is a coefficient that is also expected to reach an asymptotic constant value at very large $Re_{\tau}$.


The work here seeks to examine the applicability of the AEM to the adiabatic ASL with a focus on $\sigma_u^2$, which is needed in a plethora of applications such as footprint modelling and air quality studies \citep{banta2006turbulent,Venkatram1988}. Additionally, its numerical value is significant to wind energy assessments and to the stability of structures such as towers, bridges, and trees \citep{Lumley1965}. There is a growing number of laboratory studies supporting the logarithmic behaviour of $\sigma_u^2$ in the ISL \citep{Marusic2010,Marusic2013}. However, there have been few studies testing the logarithmic behaviour of $\sigma_u^2$ and associated coefficients in the ASL, partly because profiling $\sigma_u^2$ in the ASL demands high-fidelity measurements collected at multiple levels and obtaining such data is still challenging. Additionally, links between the logarithmic behaviour of $\sigma_u^2$ and the `-1' scaling law in the streamwise velocity energy spectrum did not receive proper attention except in a handful of studies \citep{Huang2022}.

The work here uses an extensive dataset with turbulence measurements at multiple levels within the ASL. These measurements enable a direct evaluation of the logarithmic behaviour of $\sigma_u^2$ in the near-neutral ASL and the inference of $A_1$ from the profiles of $\sigma_u^2$ as well as the one-dimensional spectra of $u'$.  Before doing so, a brief review of current formulations of $\sigma_u^2$ (or streamwise standard deviation $\sigma_u$) and the concomitant energy spectrum in turbulent boundary layers, both in the atmosphere and laboratory settings, is provided. 

\subsection{Magnitude of the turbulent velocity fluctuations}

\subsubsection{Review of ASL formulations for $\sigma_u$}
Table~\ref{tab:combined_formulations} summarizes the formulations of $\sigma_u$ and $\sigma_u^2$ found from field experiments in the ASL. According to Monin-Obkhov Similarity Theory (MOST), the streamwise velocity standard deviations can be expressed as \citep{obukhov1946turbulentnost,monin1954basic} 
\begin{equation}
    \frac{\sigma_{u}}{u_*} = \phi_{u}\left(\frac{z-d}{L}\right),
    \label{eq:sigma_alpha}
\end{equation}
where $L$ is the Obukhov length measuring the height at which mechanical production of turbulent kinetic energy (TKE) balances buoyancy production or destruction of TKE, and $\phi_{u}$ is the so-called stability correction function that varies with $(z-d)/L$. Under neutral conditions where $|L|\rightarrow\infty$, $\phi_{u}$ reduces to
\begin{equation}
    \frac{\sigma_u}{u_*} = C_0.
    \label{eq:sigma_u}
\end{equation}

\begin{table}
\begin{ruledtabular}
  \renewcommand{\arraystretch}{2}
   \begin{tabular}{llll} 
   \multicolumn{1}{c}{\bfseries Sources} & \multicolumn{1}{c}{\bfseries Formulations} & \multicolumn{1}{c}{\bfseries Fitted constants} & \multicolumn{1}{c}{\bfseries Conditions} \\ 
   \midrule 
   \text{\cite{Lumley1965}} & $\frac{\sigma_u}{u^*} = C_0$ & $C_0 = 2.1-2.9$ & \text{ASL under near-neutral conditions, pipe flow}\\
   \text{\cite{Panofsky1977}}   & $\frac{\sigma_u^2}{u_*^2} = 4 + 0.6 (\frac{\delta}{-L})^{2/3}$ &  & \text{ASL under near-convective conditions}\\
   \multirow{1}{*}{\text{\cite{Wilson2008}}} & $\frac{\sigma_u^2}{u_*^2} = [4+0.6(\frac{\delta}{-L})^{2/3}][1-(\frac{z}{\delta})^{0.38}]$ &  & \text{ASL under convective conditions} \\
   \midrule 
   \text{\cite{Townsend1976}} & $\frac{\sigma_u^2}{u_*^2} = B_1 - A_1 \text{ln} (\frac{z}{\delta})$ &  & \text{Attached eddy hypothesis and dimensional analysis}\\
   \multirow{3}{*}{\text{\cite{Perry1987}}} & \multirow{3}{*}{$\frac{\sigma_u^2}{u_*^2} = B_1 - A_1 \text{ln} (\frac{z}{\delta}) - C {Re^+}^{-1/2}$}  
              & $A_1$ = 1.03,1.26      &  \multirow{3}{*}{\text{Wind tunnel on smooth and rough walls}}  \\
           &  & $B_1$ = 2.48,2.01      &   \\
           &  & C = 6.08,7.50        &   \\
   \text{\cite{Perry1990}} & \multirow{1}{*}{$\frac{\sigma_u^2}{u_*^2} = B_1 - A_1 \text{ln} (\frac{z}{\delta}) - V[Re^+]$} 
            & $A_1$ = 1.03, $B_1$ = 2.39   & \multirow{1}{*}{\text{Wind tunnel on smooth and rough walls}}\\
   \text{\cite{Marusic1997}} 
              &\multirow{1}{*}{$\frac{\sigma_u^2}{u_*^2} = B_1 - A_1 \text{ln} (\frac{z}{\delta}) - V_g[Re^+,\frac{z}{\delta}]-W_g[\frac{z}{\delta}]$} 
              &\multirow{1}{*}{$A_1 = 1.03$, $B_1 = 2.39$}
              &\multirow{1}{*}{Wind tunnel at $Re_{\tau} \sim 734 \text{--} 13500$, ASL at $Re_{\tau} \sim 10^6$}\\
   \text{\cite{Nickels2007}} & \multirow{1}{*}{$\frac{\sigma_u^2}{u_*^2} = B_1 - A_1 \text{ln} (\frac{z}{\delta})$} & $A_1$ = 1.03, $B_1$ = 3.65 & \text{Wind tunnel at $Re_{\tau} \sim 10^4$} \\
   \text{\cite{Marusic2013}} & \multirow{2}{*}{$\frac{\sigma_u^2}{u_*^2} = B_1 - A_1 \text{ln} (\frac{z}{\delta})$} & \multirow{1}{*}{$A_1 = 1.26$, $B_1 = 2.3$}   & \multirow{1}{*}{Wind tunnel at $Re_{\tau} = 18010$}\\
            & & \multirow{1}{*}{$A_1 = 1.33$, $B_1 = 2.14$}   & \multirow{1}{*}{ASL at $Re_{\tau} \approx 628000$}\\
   \text{\cite{Samie2018}} & \multirow{1}{*}{$\frac{\sigma_u^2}{u_*^2} = B_1 - A_1 \text{ln} (\frac{z}{\delta})$} & $A_1$ = 1.26, $B_1$ = 1.95   & \text{Wind tunnel at $Re_{\tau} \sim 6000 \text{--} 20000$} \\
   \text{\cite{Huang2022}} & \multirow{1}{*}{$\frac{\sigma_u^2}{u_*^2} = B_1 - A_1 \text{ln} (\frac{z}{\delta})$} & $A_1$ = 0.91, $B_1$ = 2.25  &
          \text{ASL under neutral conditions at $Re_{\tau} = 1 \times 10^6$}\\
   \text{\cite{Hwang2022}} & \multirow{1}{*}{$\frac{\sigma_u^2}{u_*^2} \simeq B_1(Re_{\tau}) - A_1(Re_{\tau}) \text{ln} (\frac{z}{\delta})$} & $A_1$ $\sim$ 1.01 \text{--} 1.09   & \text{Wind tunnel at $Re_{\tau} \sim 6123 \text{--} 19680$}\\
   
   
    \end{tabular}
\end{ruledtabular}
\caption{\label{tab:combined_formulations}A summary of formulations for the normalized streamwise velocity variance in the ASL and based on the AEM. $\sigma_u$ represents the streamwise velocity standard deviation and $\sigma_u^2$ indicates the streamwise velocity variance. $V_g$ is a viscous correction term that depends on the viscous Reynolds number $Re^+ = z u_*/\nu$. $W_g$ is a wake correction term. The bulk Reynolds number is defined as $Re_{\tau} = \delta u_*/\nu$. }    
    
\end{table}

Several ASL measurements conducted under near-neutral conditions as well as laboratory flows have been used to fit the empirical parameter \(C_0\). \cite{Lumley1965} surveyed several experiments and stated that \(C_0\) appears independent of $z$ but varies with terrain and with values ranging between 2.1 and 2.45. The survey also noted that under varying stability conditions, vertical and horizontal velocity components exhibit distinct behaviours. While the vertical velocity standard deviation aligns well with MOST predictions, streamwise velocity components are affected by $z-d$ and by $1/L$ differently. Specifically, under near-convective conditions, a change in $z$ has a negligible effect on streamwise velocity standard deviations, whereas a change in $L$ has a pronounced effect. A recent study attributed this limitation of MOST to the anisotropy of the Reynolds stress tensor \citep{Stiperski2023}.

To account for the `non-MOST' behaviour of streamwise velocity components and the increase of streamwise wind fluctuations with the deepening of boundary layer height under near-convective conditions, \cite{Panofsky1977} proposed an empirical formulation based on ASL observations from 30 m to 87 m given by
\begin{equation}
    \frac{\sigma_u^2}{u_*^2} = 4 + 0.6 \left(\frac{\delta}{-L} \right)^{2/3}.
    \label{eq:P77}
\end{equation}
This formulation became widely cited in the atmospheric science literature \citep{Venkatram1988} although it assumed that changes in $z$ have a negligible effect on $\sigma_u$. It is to be noted that under neutral conditions Eq.~\ref{eq:P77} is similar to Eq.~\ref{eq:sigma_u}. 
Another extensive study was undertaken for unstable conditions \citep{Wilson2008} where TKE production and TKE dissipation rates were in approximate balance. These experiments reported a mild $z$-dependence of the streamwise velocity variance and a refinement to Panofsky's formulation was proposed given as
\begin{equation}
    \frac{\sigma_u^2}{u_*^2} = \left[4+0.6\left(\frac{\delta}{-L}\right)^{2/3}\right]\left[1-\left(\frac{z}{\delta}\right)^{0.38}\right].
\end{equation}
Those developments were viewed as adjustments to MOST and made no apparent contact with the AEM. In fact, after the work of \cite{kaimal1978horizontal}, \cite{bradshaw1978comments} commented that MOST formulations for the ASL appear to have missed predictions from the AEM about the role of $\delta$ in $\sigma_u^2/u_*^2$ \citep{bradshaw1967inactive}, which is briefly reviewed next.

\subsubsection{Townsend's attached-eddy hypothesis}
The AEM postulates that turbulence in wall-bounded flows can be described by a hierarchy of self-similar eddies attached to the wall \citep{Townsend1976}. These eddies are geometrically similar at different scales with their size proportional to their distance from the wall.  Using the equilibrium-layer hypothesis where the mechanical production of TKE balances the viscous dissipation, the AEM predicts a specific scaling law in the energy spectrum \citep{Perry1986,Perry1990}. Further, by integrating the streamwise velocity spectrum across all wavenumbers, $\sigma_u^2$ can be formulated as in Eq.~\ref{eqn:law_of_wall_2} for the overlap region of turbulent flows.
 This relation is a cornerstone of the AEM, linking the spectral characteristics of turbulence at a given $z$ to the profile of $\sigma_u^2$ \citep{Townsend1976, Perry1982}. 

Since the classic book by \cite{Townsend1976}, the AEM has drawn significant experimental and theoretical interest (see Table~\ref{tab:combined_formulations}). \cite{Perry1986} expanded the AEM to include near-wall regions, addressing inner flow dynamics and providing a theoretical and experimental foundation for the logarithmic law. Subsequently, \cite{Perry1987} introduced a $C {Re^+}^{-1/2}$ term to Eq~\ref{eqn:law_of_wall_2} to account for different surface types—smooth, rough, and wavy—identifying distinct coefficients ($A_1$, $B_1$, and $C$) for each. In a further advancement, \cite{Perry1990} incorporated a viscous correction term $V[Re^+]$, arguing for the formulation's independence from Reynolds number variations. Building on these foundations, \cite{Marusic1997} proposed a similarity relation to describe the streamwise velocity variance in the logarithmic region, considering inviscid attached eddies and incorporating a viscous correction term $V_g[Re^+,\frac{z}{\delta}]$ in the inner region and a wake correction term $W_g[\frac{z}{\delta}]$ in the outer region. This formulation was evaluated using wind tunnel and near-neutral ASL data with success. 

\cite{Metzger2007} tested this similarity formulation using data collected at the Surface Layer Turbulence and Environmental Science Test (SLTEST) facility, confirming the logarithmic behaviour of $\sigma_u^2$ in near-neutral ASL conditions. \cite{Nickels2007} further explored this formulation by neglecting the correction terms for the viscous and outer flow effects, which were deemed insignificant in their study. These experiments matched well with AEM predictions when setting $A_1 = 1.03$, consistent with the prior value reported in \cite{Perry1990}. 

Further experimental support for Eq.~\ref{eqn:law_of_wall_2} was provided by 
\cite{Marusic2013}, who considered four datasets including boundary layers, pipe flow and ASL measurements with $2\times 10^4 <Re_{\tau} <6\times 10^5$. Their results not only affirmed the presence of a universal logarithmic region but also estimated $A_1$ to be around 1.26 for lab flow and 1.33 for ASL data, which is different from the previous estimation of $A_1= 1.03$. \cite{Samie2018} reported an \(A_1 = 1.26\) based on wind tunnel data within the range \(6000 < Re_{\tau} < 20000\). Similarly, \cite{wang2016} found that their ASL experiments supported the estimate of \(A_1 = 1.33\), while \cite{Huang2022} used SLTEST data near the wall (just above the roughness layer) to study the high-order moments of the streamwise velocity, finding a fitted $A_1$ of about 0.9.  This value appears to be closer to the value reported by \cite{katul1998theoretical} for the near-neutral ASL derived from streamwise velocity spectra. 

Despite the variability in $A_1$, the debate continues regarding the origin of the log-law of the streamwise velocity variance. Some studies have employed spectral budget analysis and dimensional analysis, complemented by laboratory experiments \citep{Nikora1999,Banerjee2013}, to explain the logarithmic behaviour of streamwise velocity variance. Recently, \cite{Hwang2022} introduced a model inspired by Townsend's original work, addressing the spectrum in the logarithmic layer for various $z/\delta$ values. This model suggests that the coefficients $A_1$ and $B_1$ depend on the Reynolds number. Their analysis of wind tunnel data at $Re_{\tau} \sim 6123 \text{–} 19680$ indicated that the approximated $A_1$ values vary between 1.01 and 1.09. Furthermore, their model challenges the notion that a `-1' power law in the streamwise energy spectrum is necessary for the logarithmic behaviour in streamwise velocity variance. This evolving understanding underscores the need to verify the universality of the logarithmic behaviour of streamwise velocity variance, particularly within the ASL and assess to what extent the coefficient $A_1$ can be derived from the spectrum of the streamwise velocity at a single level $z$, considered next.

\subsection{Streamwise velocity energy spectrum}
In the production range, TKE is primarily generated by the mean shear and 'active' eddies. The streamwise velocity energy spectrum \(E_{uu}(k)\) as a function of streamwise wavenumber $k$ in this range typically exhibits a peak corresponding to the energy-containing eddies. In the absence of long-term trends in the record, \( E_{uu}(k) \) is expected to level off to a constant value as $k\rightarrow0$  \citep{kaimal1994atmospheric}. The AEM argues that in the limit of $Re_{\tau} \rightarrow \infty$ and when the wall-normal distance is much smaller than the boundary layer height ($z \ll \delta$), the pre-multiplied power-spectral density for $k \sim \mathcal{O}(1/\delta)$ should exhibit a characteristic $\delta$-scaling:
\begin{subequations}
\begin{equation}
  \frac{k E_{uu}(k,z)}{u_*^2} = h_1(k\delta).
  \label{eq:kF_u_delta}
\end{equation}
Similarly, for $k \sim \mathcal{O}(1/z)$, the pre-multiplied power-spectral density should follow a $z$-scaling:
\begin{equation}
  \frac{kE_{uu}(k,z)}{u_*^2} = h_2(kz).
  \label{eq:kF_u_z}
\end{equation}
\end{subequations}
In the overlap region where both the outer scaling ($\delta$) and the inner scaling ($z$) are simultaneously valid ($1/\delta \ll k \ll 1/z$), matching these scaling arguments requires \begin{equation}
  h_1(k\delta) = h_2(kz) = C_1,
  \label{eq:A}
\end{equation}
where $h_1$ and $h_2$ are two universal functions, $C_1$ is a constant independent of both $k z$ and $k \delta$, corresponding to coefficient $A_1$ in Eq~\ref{eqn:law_of_wall_2}. 

This matching also suggests that in the overlap region (or ISL), the power-spectral density must satisfy
\begin{equation}
  E_{uu}(k)=C_1 u_*^2 k^{-1},
  \label{eq:Euu_minus1}
\end{equation}
and is independent of $z$ and $\delta$. This spectrum is consistent with the presence of large, energy-containing, self-similar eddies attached to the wall \citep{bradshaw1967inactive,perry1977}.

However, observing a clear \(k^{-1}\) scaling in $E_{uu}(k,z)$ is not straightforward. It was already pointed out by \cite{antonia1993spectral} that while the \(k^{-1}\) scaling in $E_{uu}(k,z)$ was observed in their high Reynolds number wind tunnel experiments, they precluded its existence in the ASL citing unavoidable ground (and thermal) inhomogeneity and weak non-stationarities.  It was also suggested that the logarithmic layer can be influenced by large-scale motion induced by non-turbulence processes \citep{wang2016}. The presence of other turbulence structures, such as detached eddies, wake turbulence, or other flow irregularities, can contribute to the energy spectrum and obscure the \(k^{-1}\) behaviour \citep{Baars2020a} that would otherwise be observed in the production range (i.e. the range over which TKE is produced) dominated by attached eddies. 

Despite these complexities, several experiments reported a \(k^{-1}\) power law in the ASL, with \(C_1\) values ranging from 0.9 to 1.1, as summarized in Table~\ref{tab:k1_scaling}. Compared to the values in Table~\ref{tab:combined_formulations}, \(C_1\) determined from fitting \({kE_{uu}(k,z)}/{u_*^2}\) tends to be lower than \(A_1\) derived from fitting the profile of \(\sigma_u^2\). This discrepancy may be attributed to misalignment between the x-axis and the incoming wind, which could introduce biases from the cross-stream component, thereby reducing the effective production measured \citep{Huang2022}. However, no study has yet simultaneously obtained and compared both \(A_1\) and \(C_1\) from the same ASL experiments (i.e. very high Reynolds number), which partly motivated the study here.

\begin{table}

  \begin{ruledtabular}
   \begin{tabular}{lcc} 
   \multicolumn{1}{c}{\bfseries Sources} & \multicolumn{1}{c}{\boldmath{$C_1$}} & \multicolumn{1}{c}{\bfseries Conditions} \\ 
   \midrule
     \text{\cite{pond1966spectra}} & 1.0 & \text{Near-neutral ASL over sea} \\
     \text{\cite{turan1987experimental}} & 0.9, 0.92 & \text{Pipes and wind tunnels} \\
   \text{\cite{kader1991}} & 0.9 & \text{Near-neutral ASL} \\
    \text{\cite{antonia1993spectral}} & 1.0 & \text{Wind tunnel experiments over a rough surface} \\
   \text{\cite{Katul1995}} & 1.1 & \text{Near-neutral ASL over a dry lake bed} \\
   \text{\cite{katul1998theoretical}} & 1.0 & \text{Near-neutral ASL over crops and smooth wall flume} \\\text{\cite{hogstrom2002}} & $\approx 1$ & \text{Neutral ASL over grassy heath} \\
   \text{\cite{Huang2022}} & 1.01 & \text{Near-neutral ASL (SLTEST)} \\
    \end{tabular}
    \end{ruledtabular}

   \caption{A summary of experiments reporting a \(k^{-1}\) scaling in the ISL along with the corresponding \(C_1\) values.}
   \label{tab:k1_scaling}
\end{table}

\subsection{Objectives}
The logarithmic behaviour of the streamwise velocity variance and the $k^{-1}$ scaling in the energy spectra of streamwise velocity in the ASL, as well as their interconnection are to be explored. Specifically, the following questions are to be addressed:
\begin{enumerate}

\item Can a logarithmic profile of streamwise velocity variance be observed over a flat, homogeneous surface in the adiabatic ASL?
\item Using these ASL measurements, what are the dominant factors that influence the variability in  $A_1$?
\item Does a $k^{-1}$ scaling regime exist in the energy spectrum of the streamwise velocity in the adiabatic ASL with a plateau value $C_1$=$A_1$?

\end{enumerate}

To answer these questions, the paper is organized as follows: Section 2 introduces the field experiment and outlines the data processing and screening methods; Section 3 presents and discusses the results; Section 4 concludes and offers an outlook. 

\section{Methods}

\subsection{Study site}

\begin{figure}  \centerline{\includegraphics[width=0.7\textwidth]{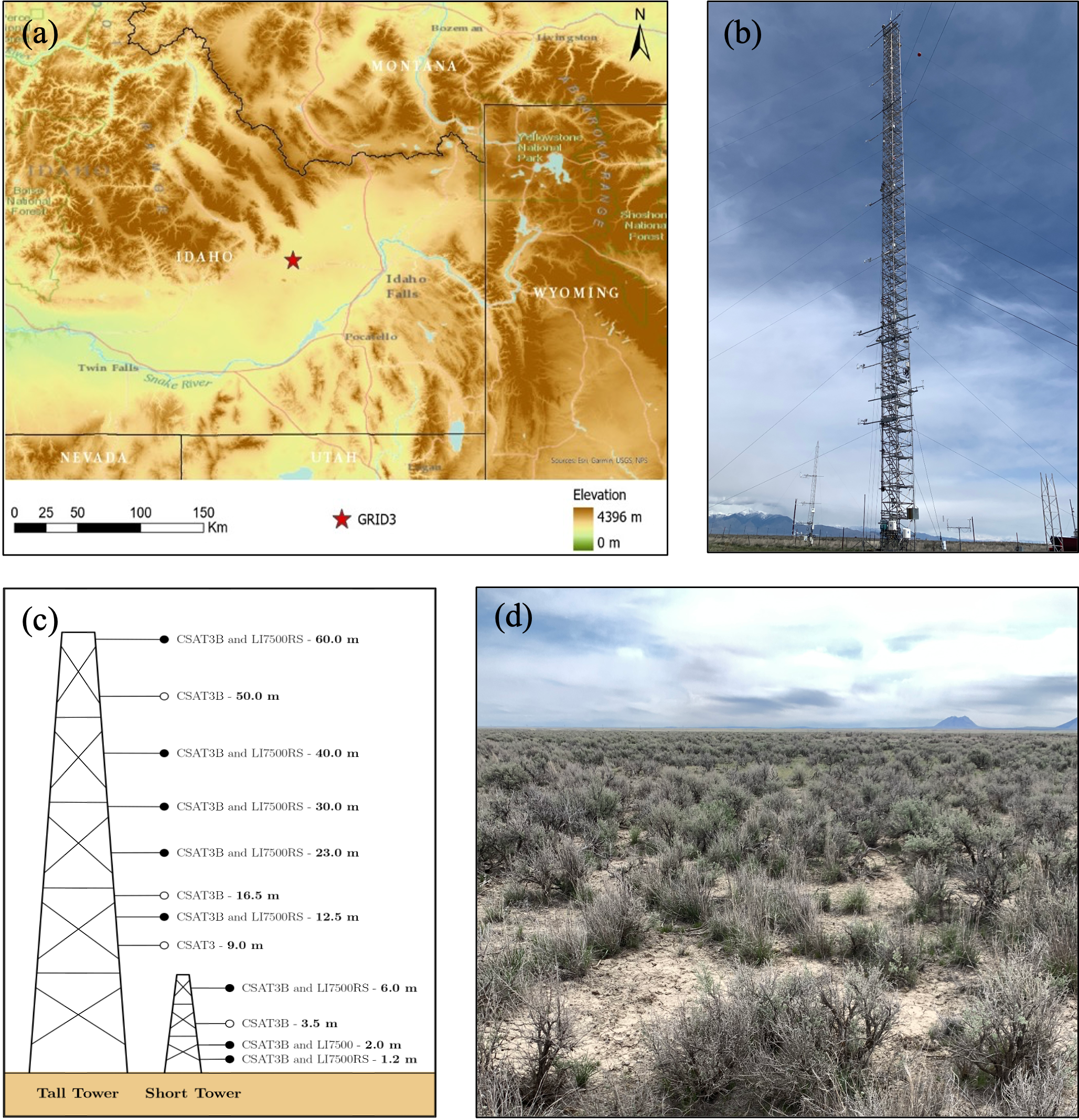}}
  \caption{(a) Topography of the Eastern Snake River Plain (ESRP), Idaho, USA. The red star marks the location of the 62-m tower GRID3. (b) The photo of the 62-m tower viewing from the northwest. (c) The configuration of the 62 and 10-m towers. (d) Dominant land surface vegetation at the site. 
  }
\label{fig:map}
\end{figure}

The study area is located in the Eastern Snake River Plain (ESRP), extending from Twin Falls, Idaho, to the Yellowstone Plateau in the northeast (see Figure~\ref{fig:map}a). The ESRP generally runs in a northeast-southwest direction and is bordered by large mountain ranges. To the northwest are the Lost River, Lemhi, and Bitterroot Mountain Ranges, which are oriented perpendicularly in a northwest-southeast direction and rise to approximately 3000 m above the mean sea level, about 1800 m above the mean elevation of the ESRP. Across the ESRP, the elevation is higher to the north and northeast and lower to the south and southwest \citep{Clawson2018}. This area is commonly influenced by general ESRP downslope winds from the northeast during the night. During the day, this area often experiences synoptic southwesterly winds, and frequent afternoon winds from the southwest that are driven by radiative heating. Under these two prevailing wind conditions, the site has a relatively flat and uniform fetch extending tens of kilometres upwind \citep{Finn2018a,Finn2018b}. Additionally, the area experiences shallow nocturnal down-valley winds from the northwest associated with the Big Lost River channel.

\subsection{Field experiments}
The experiment took place at the Idaho National Laboratory (INL) facility (43°35'30" N, 112°55'50" W). The general surface of the INL, like that of the entire ESRP, consists of rolling grasslands and sagebrush (see Figure~\ref{fig:map}d) with a $d$ close to zero \citep{Finn2016,Clawson2018}, yielding a median $z_0$ of approximately 3 cm for southwest winds and 3.8 cm for northeast winds. From September 20, 2020 to April 22, 2021, 12 measurement levels of eddy-covariance systems were deployed on two towers. The first is a 62-m meteorological tower (GRID3) instrumented at eight levels (9, 12.5, 16.5, 23, 30, 40, 50, and 60 m), while the second is an auxiliary 10-m tower instrumented at four levels (1.2, 2, 3.5, and 6 m) (see Figure ~\ref{fig:map}b,c). Despite the relatively flat and homogeneous surface, this setup results in different measurement footprints between the two towers.

The eddy-covariance systems used in the experiment include two models of triaxial sonic anemometers (CSAT3B and CSAT3, Campbell Scientific, Inc.) and two models of infrared gas analyzers (IRGA; LI7500RS, and LI7500, LICOR, Inc.). The sonic anemometers measure the velocity components along the north-south, east-west, and vertical directions, respectively, relative to the reference frame fixed to the sonic anemometer. Sonic azimuth was estimated by comparing the calculated and measured wind direction from sonics and wind vanes. It was found that the sonic anemometers were oriented slightly toward the north-northwest, rather than directly north. The IRGAs measure the densities of water vapour and \(\mathrm{CO_2}\).

The 62-m tower is equipped with retractable square booms (Tower Systems, Inc) measuring 3.6 metres (12 ft), mounted horizontally to provide stable platforms for the sensors. On the 10-m tower, 1.8-m (6 ft) poles were used, ensuring that the sensors were placed approximately 1.5 m away from the tower’s structure. This study focuses primarily on data collected from the 62-m tower. Given that all sonic anemometers were positioned at least 3 m from the tower, flow distortion is considered minimal and Angle of Attack corrections were not applied.

The sampling frequency was set to 10 Hz for all the anemometers. The CSAT3s have vertical sonic paths of 10 cm and horizontal paths of 5.8 cm, operating in a pulsed acoustic mode. Corrections for the effects of humidity and density fluctuations on the turbulent fluctuations of sonic temperature and the densities of water vapour and \(\mathrm{CO_2}\) are detailed elsewhere \citep{gao2024}.  These scalar measurements were only needed to estimate whether $L$ is sufficiently large and thus whether the flow is near-neutral.


\subsection{Data processing}

\begin{figure}
    \centerline{\includegraphics[width=0.7\textwidth]{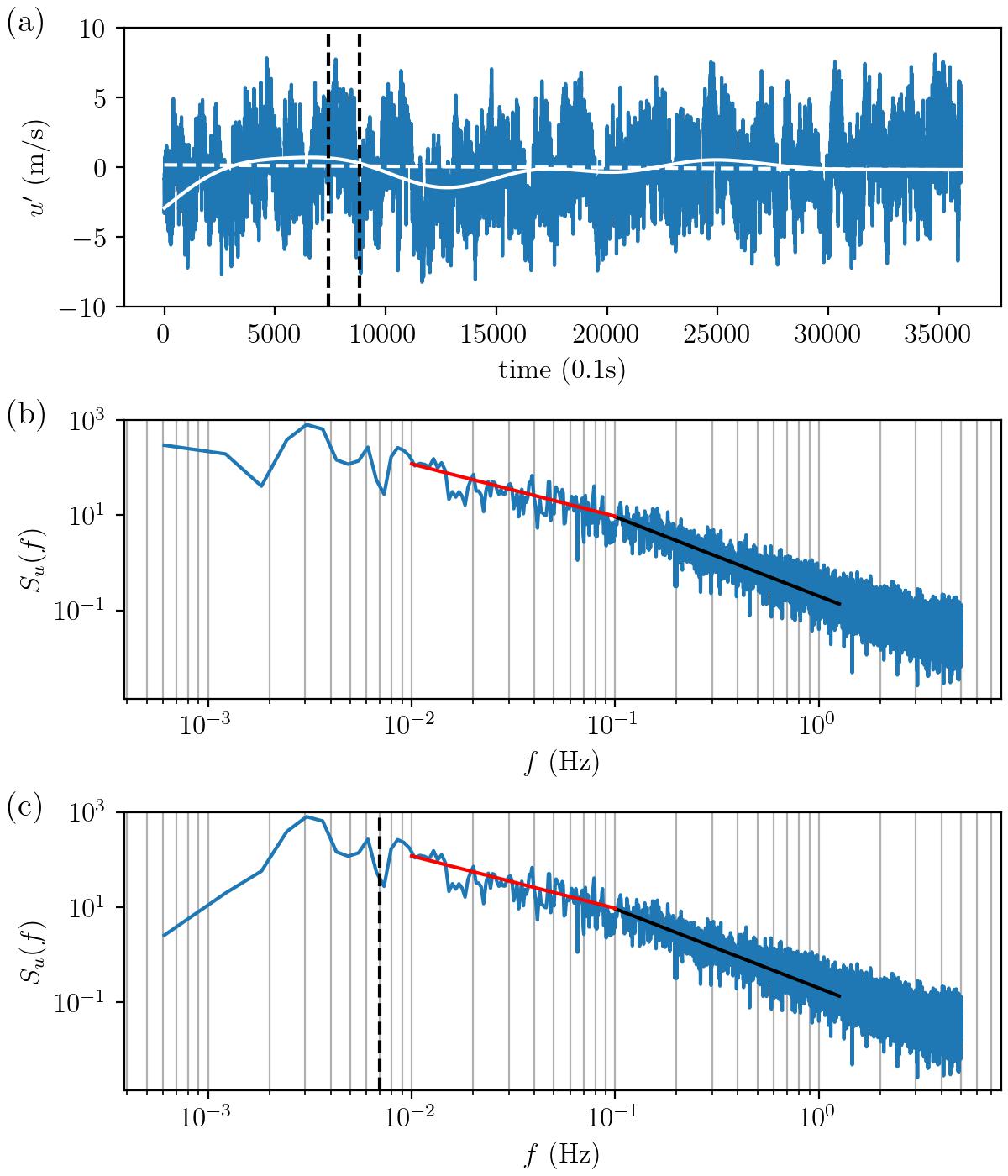}}
    \caption{Streamwise velocity time series, $u$, sampled at 10 Hz from the triaxial sonic anemometer at 12.5 m on Sep. 25, 2020, at 15:00 local time. 
    (a) Turbulent component of streamwise velocity after coordinate rotation: the \textit{dashed white line} shows the linear fit, the \textit{solid white line} represents high-pass filtered low-frequency signals that are non-linear, and the \textit{vertical dashed black lines} mark the filter size for the high-pass method. (b) Energy spectrum of $u$ after linear de-trending: \textit{solid red line} and \textit{solid black line} represent $f^{-1}$ and $f^{-5/3}$ scalings, respectively. (c) The energy spectrum of $u$ after high-pass filtering: the \textit{vertical dashed black line} indicates the cut-off frequency corresponding to a 2000-m wavelength, for which the filter size is 142 seconds.}
    \label{fig:processing}
\end{figure}
High-frequency spectral corrections are neglected because the contribution from the high-frequency region to turbulence intensity is small. For example, the data indicate that the errors caused by path averaging are 2-3 orders of magnitude smaller than the overall variance. Data processing involves coordinate rotation, de-trending, and neutral case screening. When implementing certain constraints on these data processing steps, the total of 4788 hours was reduced to 120 hours as described next. 


\subsubsection{Coordinate rotation}
Assuming the sonic anemometers are nearly level, requiring only minor tilt corrections to align the vertical axis perpendicular to the mean flow, the double-rotation method \citep{wilczak2001} involves two sequential rotations: first, a pitch rotation to set the mean vertical velocity component ($\overline{w}$) to zero, and second, a yaw rotation to set the mean lateral velocity component ($\overline{v}$) to zero. Given the relatively flat and homogeneous terrain, this standard method was initially applied to correct the tilt of the anemometers. 

Since achieving perfectly levelled sonic anemometers on the 62-meter tower is challenging, the planar fit method \citep{wilczak2001} was applied to assess the consistency of turbulent quantities between the two coordinate rotation methods. This method, which is more robust to initial misalignments, fits a plane to the wind data over an extended period, based on specific wind direction sectors. 



\subsubsection{Averaging period}
In the analysis of ASL data, fluctuations with periods less than about one hour are generally considered turbulence, while slower fluctuations are synoptic-scale and treated as part of the mean flow \citep{wyngaard1992atmospheric}. Consequently, a one-hour averaging period was selected as a compromise between the need to resolve large eddies reliably and stationarity considerations. However, \cite{Metzger2007} suggests that the neutral periods are often short, on the order of several minutes. Previous studies have used various averaging periods depending on the stability and conditions of the boundary layer. For instance, 10 minutes were used in the stable boundary layer in the CASES-99 experiment \citep{banta2006turbulent}, 30 minutes in the unstable conditions when mechanical production of TKE was compared to viscous dissipation \citep{Wilson2008}, 20 minutes in neutral conditions in the SLTEST experiment \citep{Metzger2007}, and 1 hour in neutral conditions in the same SLTEST experiment \citep{Hutchins2012}. Additionally, 30 minutes were used in SLTEST \citep{Huang2022}, 15 minutes in the unstable LATEX experiment \citep{Li2011}, and 1-minute for stable and 30 minutes for unstable conditions across 13 datasets \citep{Stiperski2023}.

A shorter sampling period, while increasing steadiness in mean meteorological conditions, can also affect statistical convergence by reducing sample size, increasing sensitivity to outliers, and potentially biasing the representative nature of the data. In general, the difference between ensemble-averaging (the sought quantity) and time-averaging is labelled as systematic bias.  This bias declines with reduced $2 T_L/T_p$ \citep{lenschow1994long}, where $T_L$ is the integral time scale of a flow variable and $T_p$ is the averaging period. Typical $T_L$ values for the streamwise velocity are on the order of 1-2 minutes and hence selecting $T_p=60$ minutes is acceptable for reducing the systematic bias.
However, potential contamination by trends and changes in the meteorological conditions to the near-neutral ASL turbulence is acknowledged, and some trend removal must still be performed to remove synoptic weather phenomena.

\subsubsection{De-trending}
After the coordinate rotation, the time series in Figure~\ref{fig:processing}a shows a superposition of the high-frequency turbulent fluctuations and low-frequency oscillations. De-trending is employed to remove long-term trends or patterns in the data, which can result from various factors such as instrument drift and atmospheric changes \citep{moncrieff2004}. By removing the low-frequency content, the analysis can focus on short-term variability (i.e., turbulence) and ensure a certain amount of stationarity. 

A common method in micrometeorology is linear detrending, where the line of best fit over the 1-hour averaging period (dashed white line in Figure~\ref{fig:processing}b) is subtracted from the original time series. While primarily affecting the low-frequency part of the signal, linear de-trending can impact all frequencies and introduce oscillations at higher frequencies \citep{moncrieff2004}. 

Alternatively, high-pass filtering can isolate low-frequency signals by convolving the original data with a transfer function in the frequency domain, a method often used in post-processing  \citep{Nickels2005,Hutchins2012,Metzger2007}. However, determining a clear cut-off frequency to separate turbulent motions from long-term trends is challenging, posing a risk of filtering out low-wavenumber turbulent motions. A $10^{th}$ order Butterworth filter was applied in this work with a cut-off frequency equivalent to a 2000-m cut-off wavelength. It translates to a filter size of 100 to 350 seconds that varies with mean wind speed at different levels. Those sizes align closely with the observed time period of low-frequency signals in the time series (indicated by the two vertical lines in Figure~\ref{fig:processing}b). This filter size is consistent with previous turbulent boundary layer studies that selected 180 seconds in \cite{Hutchins2012} and 181.8 seconds in \cite{Puccioni2023}. 

As shown in Figure~\ref{fig:processing}c, the linear detrended streamwise velocity energy spectrum $S_u(f)$ levels off as $f \to 0$. In contrast (see Figure~\ref{fig:processing}d), the high-pass filtered $S_u(f)$ shows more removal of low-frequency content (very large scale motions) than linear de-trending, with a much faster decay as $f \to 0$ as expected. 
Nevertheless, the best method depends on site conditions. Therefore, both methods are initially applied, and a comparison is performed to ensure the derived AEM coefficients are not sensitive to the de-trending approach.

\subsubsection{Turbulence statistics}
After de-trending, hourly averaging of all instantaneous data are conducted to obtain first the mean component and then the turbulent fluctuations as $u^\prime = u-\overline{u}$, $v^\prime = v-\overline{v}$, $w^\prime = w-\overline{w}$, and $T^\prime = T-\overline{T}$, where $T$ is the air temperature. The Reynolds stress methods and profile methods are two conventional approaches to estimating the friction velocity in the ASL and those two methods will also be compared for reference.  
\begin{enumerate}
    \item {\textbf{Reynolds stress method}}
    In an ideal ASL that is high Reynolds number, stationary, planar homogeneous, lacking subsidence, and characterized by a zero mean pressure gradient, the momentum fluxes must be invariant to variations in $z$.  To test the constant flux layer assumption and determine $u_*$ to be used as a normalizing velocity in the AEM, a local friction velocity is defined as
\begin{equation}
   u_* =[\overline{u^\prime w^\prime}^2+\overline{v^\prime w^\prime}^2]^{1/4}.
   \label{u_ec}
\end{equation} 
\item {\textbf{Profile method}}
Another approach to estimating $u_*$ is based on the logarithmic law of the mean velocity as shown in Eq.~\ref{eqn:law_of_wall_1}.
Measurements of the mean velocity profile can be used to estimate the friction velocity by fitting the measured $\overline{u}$ against $\ln(z)$. To do so, it is necessary to identify the beginning and ending points that follow a log-linear relation, which can involve some subjectivity. The magnitude of the friction velocity is directly related to the choice of the von K\'arm\'an constant \citep{kendall2008method}. 
\end{enumerate}

Agreement between these two methods is also used when identifying the span of the ASL.
\subsection{Data selection}
\begin{figure}    \centerline{\includegraphics[width=\textwidth]{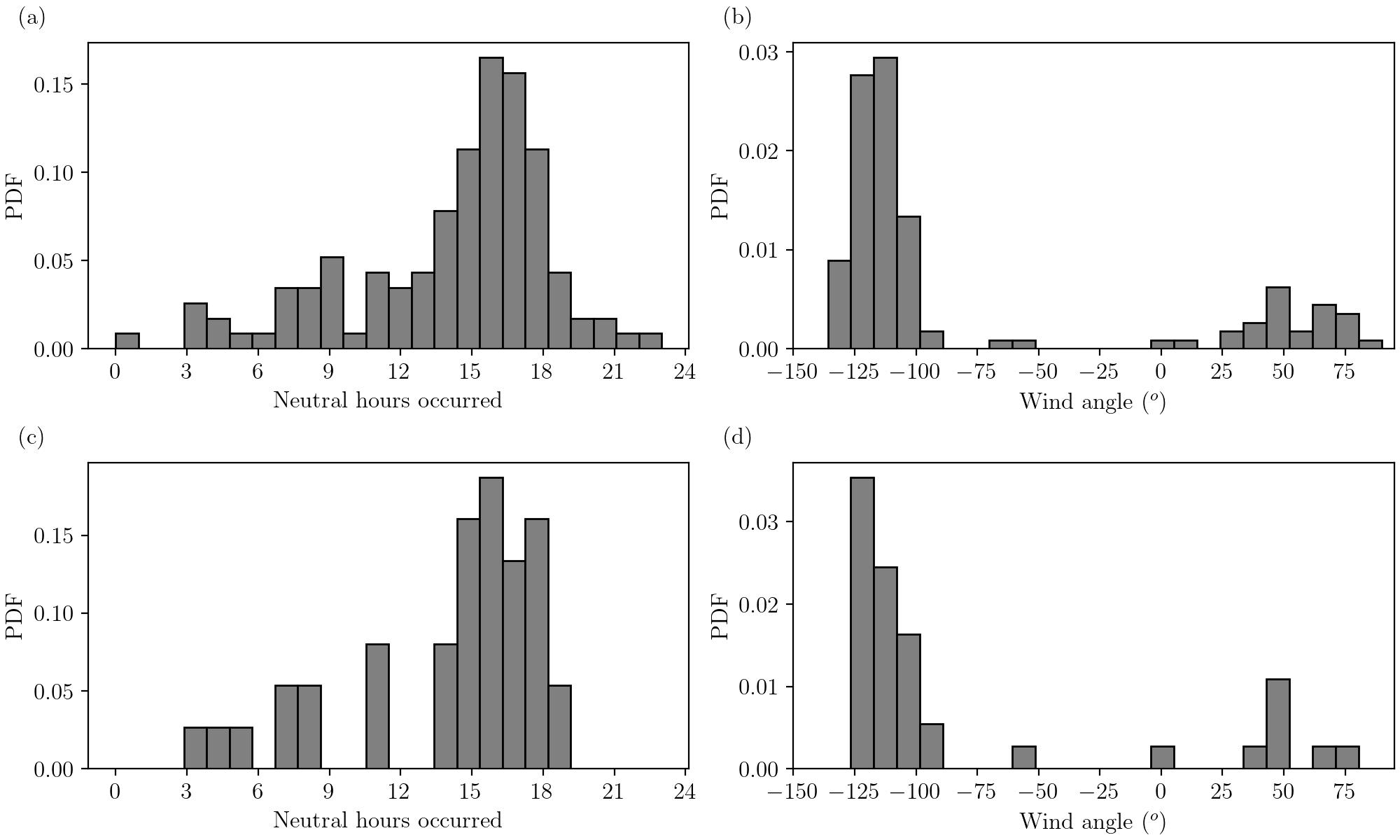}}
    \caption{Probability density functions of the 120 near-neutral cases (a) by time of day and (b) by wind angles at 16.5 m relative to true north, with positive values denoting clockwise, negative values counterclockwise, and 0 denoting northerly wind. Neutral conditions are identified during hours when $|z/L|<0.1$ is satisfied. Probability density functions of the 39 post-screening cases with $Re_{\tau}>4 \times 10^7$ and $R^2>0.6$: (c) by time of day and (d) by wind angles.}
    \label{fig:neutral_hist}
\end{figure}
From the 210 days of data recorded, the focus is placed on a subset that is approximately neutrally stratified. The stability of the surface layer is characterized by the stability parameter $|z/L|$ ($<0.1$ for near-neutral conditions). Since multiple levels are considered, in the calculation of stability parameter $z$ is represented by the geometric mean of the chosen levels, and $L$ is estimated by the median value of the Obukhov length ($L = -\overline{T}u_*^3/\kappa g \overline{w'T'}$) for these levels, where $g$ is the gravitational acceleration and the hourly-averaged sonic temperature $\overline{T}$ is assumed to be a good approximation of the virtual potential temperature $\theta_v$. This step resulted in 142 cases remaining.  The following criteria are used for further data selection to be consistent with previous ASL studies \citep{Li2011}:
\begin{enumerate}
\item Wind must face the sonic anemometers. The angles between the hourly-mean wind direction and the sonic anemometers have to be smaller than 120$\degree$ such that the interference and data contamination from the anemometer arms, tripods
and other supporting structures are minimized, which further narrowed down the runs to 139;
\item The turbulence intensity ($\sigma_u/\overline{u}$) must be less than 0.5 to justify the use of Taylor’s frozen-turbulence hypothesis \citep{taylor1938}, leading to 134 cases remaining;
\end{enumerate}
After applying the above screening procedures and removing cases with pressure measurement errors associated with the gas analyzer, 120 hours of data are retained for further analysis. Figure \ref{fig:neutral_hist}a illustrates the probability density function (PDF) of near-neutral conditions, showing that neutrally stratified ASL can occur throughout the day, peaking at around 16:00 local time. The wind is consistently dominated by a west-southwesterly wind (as shown in Fig.~\ref{fig:neutral_hist}b), with few instances of wind from the east-northeast, likely occurring at night.

\section{Results}
\subsection{Determination of the inertial sublayer}

\begin{figure}
\centerline{\includegraphics[width=0.9\textwidth]{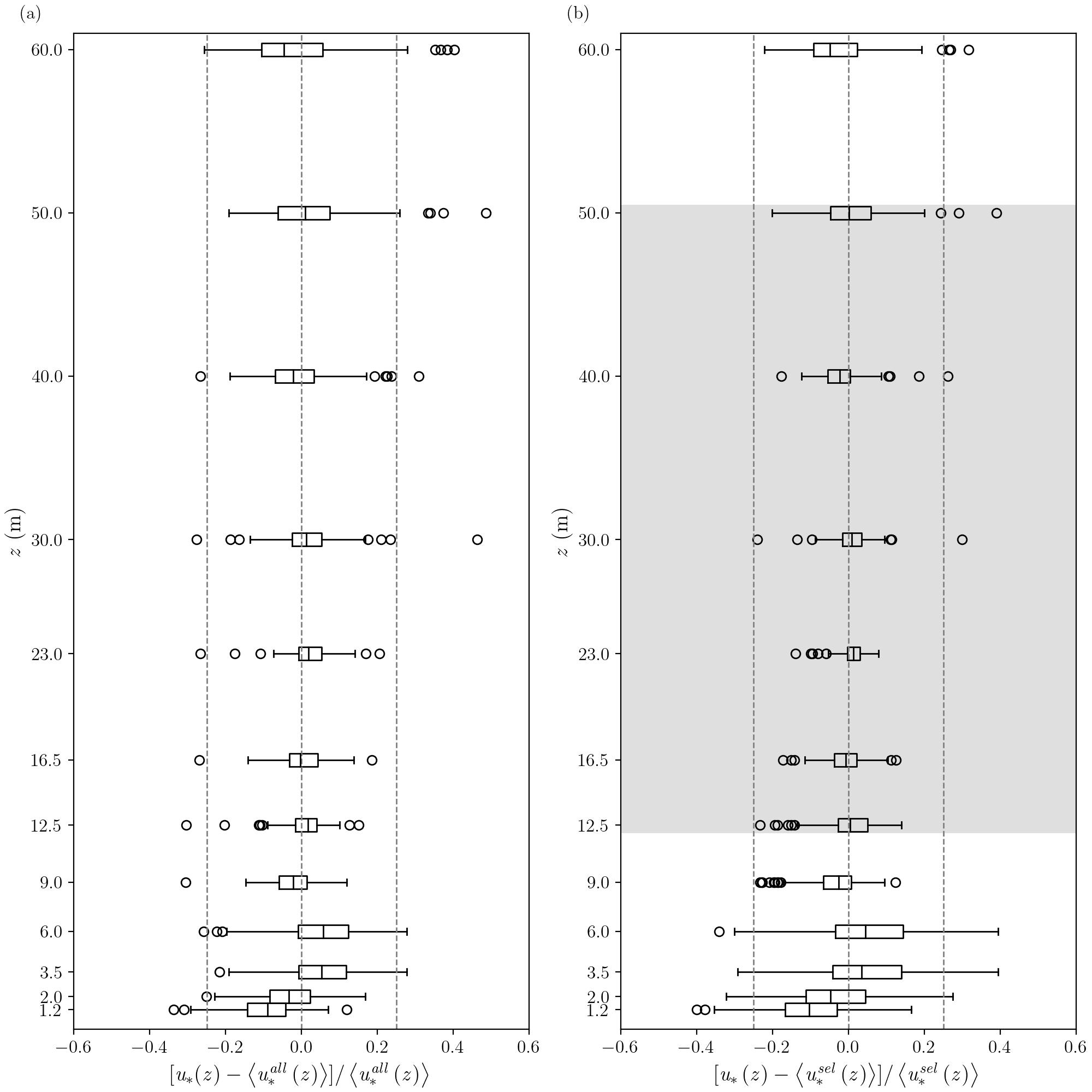}}
\caption{Deviation of locally measured friction velocity from the vertical mean averaged across (a) all twelve levels and (b) the six levels within the ISL for the linear de-trended measurements. Box plot interpretation: From left to right, lines signify the minimum, first quartile, median, third quartile, and maximum values. \textit{Black open circles} denote outliers. \textit{Vertical dashed lines} are set at -0.25, 0, and 0.25. The \textit{grey shaded} region highlights the ISL identified between 12.5 metres and 50 metres, which is the operational range used in evaluating the AEM. The terms $
\left<u_*^{all}(z)\right>$ and $\left<u_*^{sel}(z)\right>$ represent the friction velocity averaged over all twelve levels and the six shaded levels, respectively.}
\label{fig:u*_ldtr_filt_vert_6l}
\end{figure}

\begin{figure}
\centerline{\includegraphics[width=\textwidth]{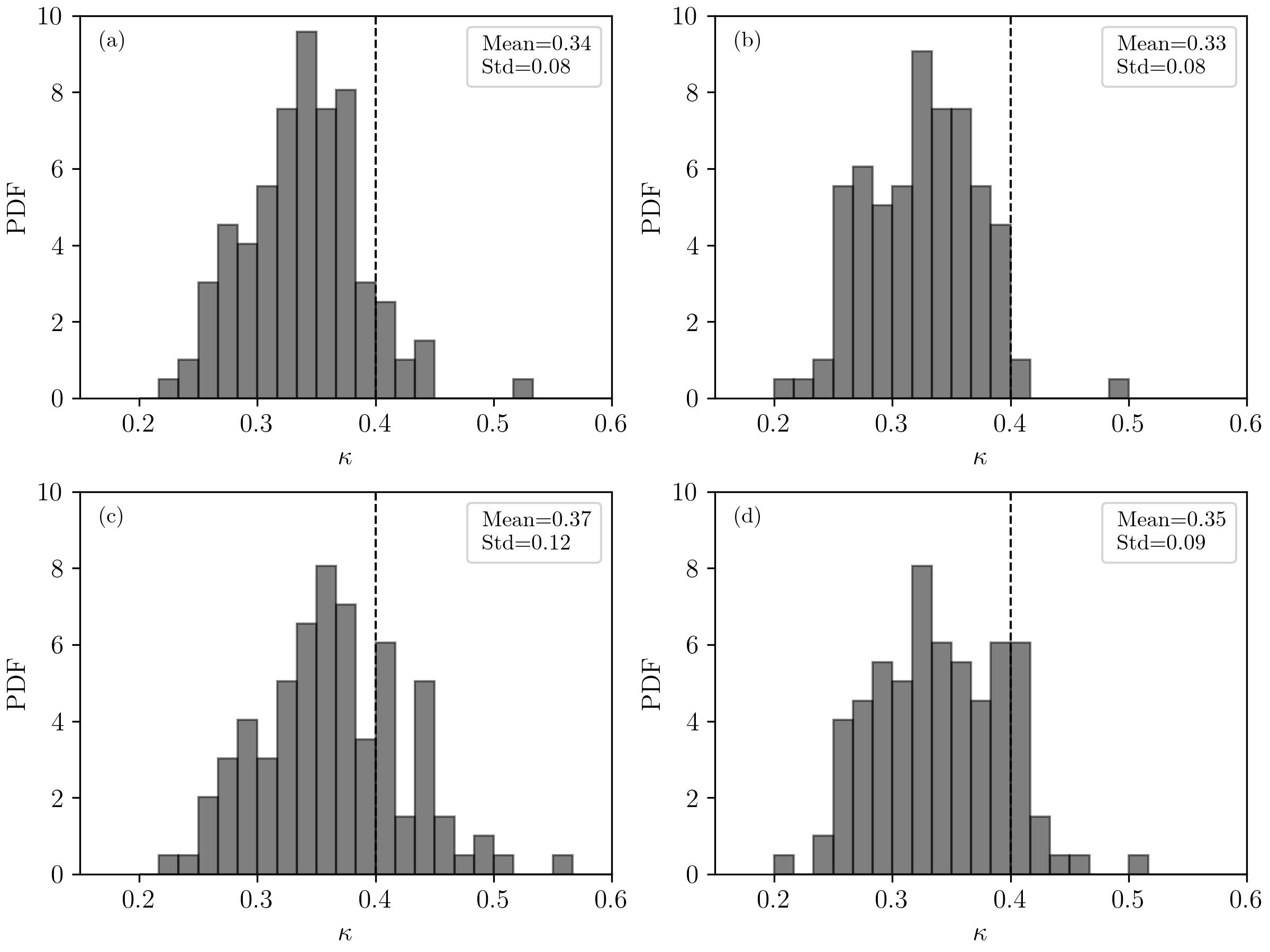}}
\caption{Probability density functions for the inferred $\kappa$ values using the mean horizontal velocity from the following levels: a) 9 to 50 metres; (b) 9 to 60 metres; (c) 12.5 to 50 metres; (d) 12.5 to 60 metres. \textit{Vertical dashed lines} indicate $\kappa$ equals 0.4. Insets show the computed mean and standard deviation of the fitted $\kappa$.}
\label{fig:kappa_test}
\end{figure}
The local friction velocity computed using Eq.~\ref{u_ec} is first investigated. Figure~\ref{fig:u*_ldtr_filt_vert_6l}a illustrates the vertical distribution of local $u_*(z)$ and their deviations from the vertical average across all 12 levels ($\left<u^{all}_*\left(z\right)\right>$, where the angle bracket indicates vertical average). Notably, the friction velocity at the lowermost four levels exhibits significant deviations from the vertical average with increasing trend in $z$. Three potential explanations are considered for this observation. First, the lower four levels are mounted on a separate 10-m tower, located approximately 10 metres from the 62-m tower, as shown in Figure~\ref{fig:map}b. Despite the flat and aerodynamically homogeneous surface cover, variations in tower placement may influence turbulent momentum flux measurements near the ground. Second, although the sagebrush is relatively short (tens of centimetres) and the lowest anemometer level is at 1.2 metres, the $z_0$ derived from the log mean profile has a median value of around 0.07 metres. According to \citet{garratt1994atmospheric}, the roughness sublayer's height can be 10 to 150 times $z_0$. Assuming a factor of 100 times the roughness length (i.e., 7 m), the roughness sublayer would extend just above the 6 m level. Last, volume averaging by the sonic anemometer path length (10-15 cm) has a disproportionate impact on the levels close to the ground - where the entire inertial subrange at finer scales is `filtered' out by instrument averaging,  leading to an underestimation of turbulence intensity (and thus, the friction velocity) at those levels. Therefore, the analysis here excludes the bottom four levels when defining the ISL.

The remaining 8 levels, from 9 metres to 60 metres, show consistent local $u_*$ vertical distribution. To test whether all these 8 levels are within the ISL, Eq.~\ref{eqn:law_of_wall_1} is further applied to fit a friction velocity $u_*^{fit}$, assuming a $\kappa$ of 0.4. 
The $u_*^{fit}$ is consistently 10-20\% higher than the local $u_*(z)$ at these 8 levels, implying that the measured turbulent momentum flux and mean velocity at these levels are not entirely consistent with a logarithmic mean velocity profile with a $\kappa$ of 0.4. 

To identify data consistent with a logarithmic mean velocity profile with a $\kappa$ close to 0.4, the local $u_*$ is used to fit $\kappa$. Using the remaining 8 levels (ranging from 9 to 60 metres), four scenarios for fitting the $\kappa$ are tested: (a) using data from 9 to 50 metres; (b) using data from 9 to 60 metres; (c) using data from 12.5 to 50 metres; and (d) using data from 12.5 to 60 metres. The PDF of the inferred $\kappa$ values is shown in Figure~\ref{fig:kappa_test} for all these cases. The results indicate that only when using data from 12.5 to 50 m does the fitted $\kappa$ have a mean close to 0.4 and exhibit a quasi-Gaussian distribution in the spread, while other scenarios produce $\kappa$ values that either deviate largely from 0.4 or do not follow a Gaussian distribution in their spread. The same analysis performed on high-pass filtered data produces results similar to those from the linearly de-trended data and are thus not presented. 

Based on this analysis of local $u_*$, the data between 12.5 metres and 50 metres are selected for further analysis of the AEM, and this range is treated as ISL (with the deviation of the locally measured $u_*$ from their vertical average shown in Figure~\ref{fig:u*_ldtr_filt_vert_6l}b). The friction velocity used for normalization in the following analysis is the vertically averaged friction velocity across this range ($\left<u^{sel}_*\left(z\right)\right>$). 

\subsection{Magnitude of the turbulent velocity fluctuations}

\begin{figure}
\centerline{\includegraphics[width=\textwidth]{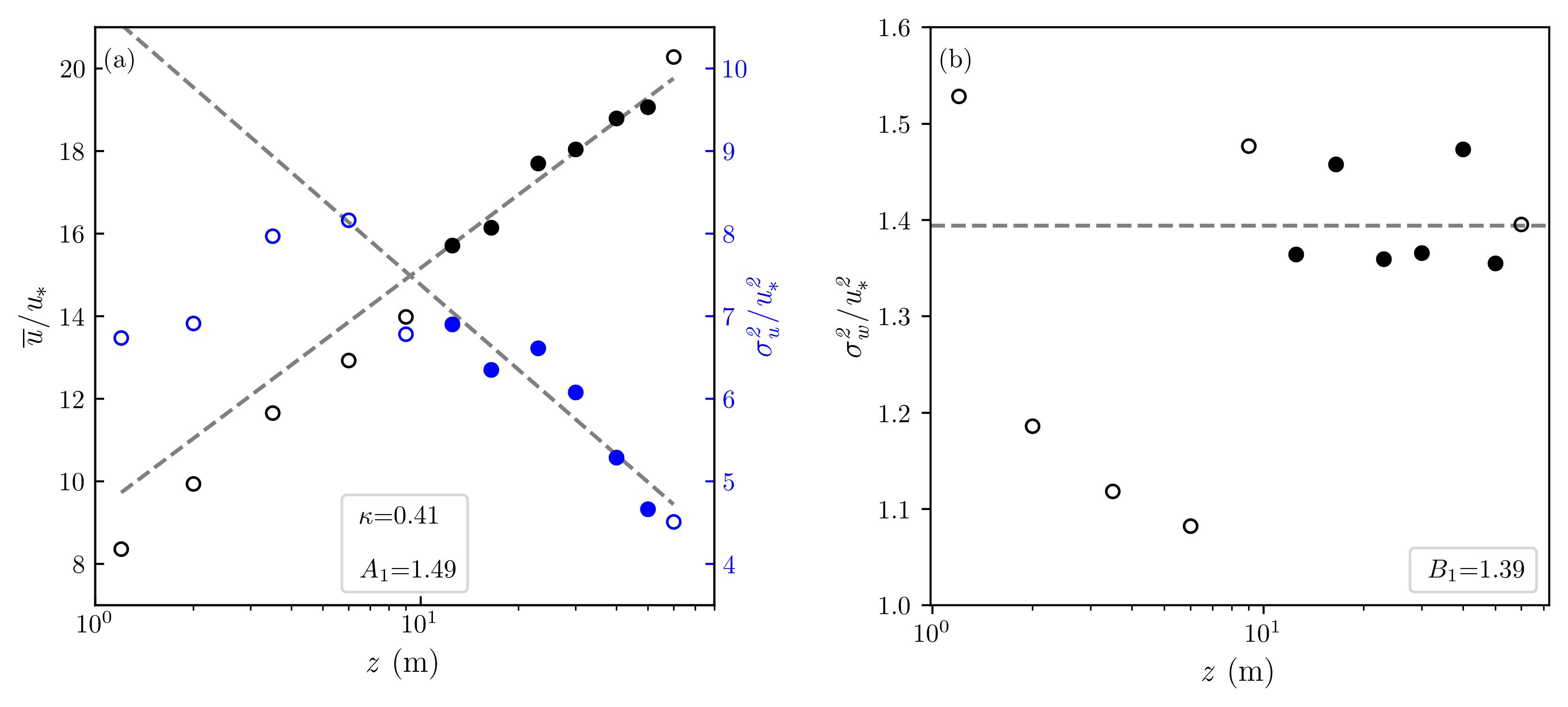}}
\caption{Fitting $\kappa$, $A_1$ and $B_2$ with high-pass filtered data on Sep. 25, 2020, at 15:00 local time. (a) The mean velocity profile (in black), and the streamwise velocity variance profile (in blue). (b) The vertical velocity variance profile.
\textit{Closed circles} indicate the selected data within the ISL. \textit{Dashed grey lines} denote the linear regression based on the selected data within the ISL. Insets are the results of the fitting.}
\label{fig:profile}
\end{figure}

\subsubsection{Estimation of $\kappa$, $A_1$ and $B_2$ in ASL}
With the $u_*$ and $\sigma_u^2$ within the ISL measured, a linear regression is conducted to derive the values of $A_1$ by fitting the $\sigma_u^2$ to $\ln(z)$ utilizing Eq.~\ref{eqn:law_of_wall_2}. Notably, this fitting method for obtaining $A_1$ does not require direct measurement of $\delta$ but cannot be used to infer the $B_1$ constant of the AEM. Figure~\ref{fig:profile}a demonstrates one such linear regression for the hour of 1500LT on Sep. 25, 2020, detrended by high pass filtering. The logarithmic mean velocity profile is also shown for comparison. The results indicate that the fitted $\kappa$ (=0.41) is very close to the accepted 0.4 value. Moreover, the $\sigma_u^2$ profile aligns with AEM predictions confirming the logarithmic dependence on $z$ of streamwise velocity variance within the ISL (highlighted by closed circles) at a very high Reynolds number.  

To estimate $B_2$ in Eq.~\ref{eqn:law_of_wall_3}, $\sigma_w^2$ is fitted against the local $u_*^2$ within the ISL, and the resulting slope is retrieved.
If the vertical velocity energy spectrum is approximated by two regions: a flat region from $k=0$ to $k=k_a$ (energy splashing) and an inertial subrange region from $k=k_a$ to $k=\infty$, then the area under this idealized spectrum with $k_a=(\kappa z)^{-1}$ and TKE production balancing TKE dissipation is $(\sigma_w/u_*)^2=(5/2) C_{o,w}$. Here, $C_{o,w}$ is the Kolmogorov constant for the one-dimensional vertical velocity spectrum in the inertial subrange (=0.65). This $B_2$ equals 1.65 (instead of 1.39 inferred here, see Figure~\ref{fig:profile}b) is thus anticipated for infinite Reynolds number when the transition wavenumber $k_a=(\kappa z)^{-1}$ and the energy splashing region due to wall effects extends all the way from $k_a$ to $k=0$. In the area calculations of the vertical velocity spectrum, it is interesting to note that $(3/2) C_{o,w}$ originate from extending the inertial subrange from $k=k_a$ to $k=\infty$ whereas only $C_{o,w}$ is the contribution from the large eddies (i.e. eddies exceeding $\kappa z$). Thus, a plausibility argument to the reduced $B_2$ measured here is that the 10 Hz sampling frequency and anemometer path averaging may disproportionately under-resolve the inertial subrange components of \( w \) compared to its longitudinal velocity counterpart. Nonetheless, similar values for $B_2$ (i.e., around 1.4) have been observed in ASL literature \citep{yang2018}.

\begin{table}
    \centering
    \begin{ruledtabular}
    \begin{tabular}{lccccccccc}
    & \multicolumn{3}{c}{\bfseries \boldmath{$\kappa$}} & \multicolumn{3}{c}{\bfseries \boldmath{$A_1$}} & \multicolumn{3}{c}{\bfseries \boldmath{$B_2$}}\\ 
    \cmidrule(r){2-4} \cmidrule(r){5-7}  \cmidrule(r){8-10} 
    & I & II & III &  I & II & III & I & II & III\\
    \midrule
    \text{Mean}  &0.38 &0.37 &0.38 &0.18 &1.25 &1.22 &1.37 &1.45 &1.40\\
    \text{Std} &0.07 &0.12 &0.05 &1.50 &0.84 &0.33 &0.22 & 0.18 &0.11 \\
    \midrule
    \text{Mean}  &0.39 &0.37 &0.38 &0.33 &1.15 &1.24 &1.31 &1.43 &1.40\\
    \text{Std} &0.08 &0.07 &0.04 &1.24 &0.82 &0.39 &0.25 & 0.19 &0.11 \\
    \end{tabular}
    \end{ruledtabular}
    \caption{Comparison of the fitted values of the von Kármán constant (\(\kappa\)), AEM coefficient (\(A_1\)), and \(B_2\) between Group I (linear-detrended data), Group II (high-pass filtered data), and Group III (high-pass filtered data with \(Re_{\tau} > 4 \times 10^7\) and \(R^2 > 0.6\)). \(B_2\) is estimated from the slope of \(\sigma_w^2\) over \(u_*^2\) within the ISL. The \(R^2\) values represent the coefficient of determination of the fits. The top two rows display results from double rotation, while the bottom two are from planar fit.}
    \label{tab:kappa_A1}
\end{table}

To assess the consistency of the logarithmic behaviour of the streamwise velocity and associated AEM coefficients in the near-neutral ASL, the same fitting procedures are applied across all 120 neutral cases. The comparison of the results between linearly detrended data and high-pass filtered data can be found in Table~\ref{tab:kappa_A1}. The analysis shows that the estimated values of $\kappa$ for both methods are similar, yet smaller than 0.4, aligning with previous experiments \citep{Andreas2006,Marusic2010} as well as other theories based on the co-spectral budget \citep{katul2013co} where $\kappa$ was linked to the Rotta constant, an isotropization constant linked to the rapid distortion theory, and the Kolmogorov constant. Furthermore, the fitted $B_2$ are relatively consistent across two detrending methods. As a result, the focus of the current analysis is on $A_1$. The high-pass filtered data exhibit a more robust logarithmic scaling of $\sigma_ u^2$ than the linearly detrended data. This outcome is based on the finding that the linear detrending method yields a wider range of $R^2$ values with an average of 0.67 and a standard deviation of 0.32. In contrast, the high-pass filtering shows both a higher mean (0.81) and a reduced standard deviation (0.23) in $R^2$ scores. Moreover, the derived $A_1$ values of linear-detrended data exhibit significant variability, often departing considerably from the conventional range of 1 to 1.3, and frequently resulting in negative $A_1$ values (31.63\% and 9.18\% negative occurrence for linear detrending and high-pass filtering, respectively). The potential reasons for this deviation are discussed next.

Given an averaging period of 1 hour, the turbulence data might experience unsteadiness associated with larger-scale meteorological influences such as static pressure gradients. According to \cite{hogstrom2002}, `inactive' turbulence becomes more significant as the pressure gradient increases, potentially obscuring the scaling behaviour predicted by the AEM. While the linear-detrending method only removes the mean trend, it might not sufficiently eliminate the (non-linear) large-scale meteorological influences, potentially leading to negative fitted values of \(A_1\). This underscores the importance of removing larger-scale meteorological influences in ASL data when comparing them to a canonical turbulent boundary layer. Furthermore, sensitivity tests indicate that the fitting outcomes are not sensitive to the choice of the cut-off frequency in the 100-200 second range for the high-pass filtering method. Consequently, the following analysis focuses on high-pass filtered data.

Finally, Table~\ref{tab:kappa_A1} presents results from both the double rotation and planar fit methods, showing no substantial differences between them except for the fitted \(A_1\), which may be more sensitive to coordinate transformations due to its higher moment nature. As shall be seen in Section \ref{section}, once data quality control is applied, the difference between the double rotation and planar fit methods becomes much small even for fitted $A_1$. For simplicity, the following discussion is primarily based on the double rotation method. 

\subsubsection{Objective data quality control} \label{section}
To understand the variations in the fitted $A_1$, key factors that influence data quality are investigated including statistical fitting performance, non-stationarity effects, the presence of a constant $\sigma_w$ with height in the ISL, and the Reynolds number effect.
\begin{enumerate}
    \item{\textbf{Fitting Performance}} 
Before discussing the fitted coefficients of the AEM to the data here, it is necessary to evaluate the goodness of the logarithmic scaling for $\sigma_u^2$. The coefficient of determination (\(R^2\)) from the logarithmic fitting is examined and anomalous cases with \(R^2\) less than 0.6 are filtered out to ensure that only high-quality fits are considered when evaluating $A_1$.

\item{\textbf{Non-Stationarity Effects}}
Stationarity is a necessary condition for the AEM and for the identification of the ISL. Two indices of stationarity (IST) are defined here as:

\begin{equation}
    \mathrm{IST_{wspd}} = \frac{|\frac{1}{12} \sum_{i=1}^{12} {\sigma_u^i}^2-{\sigma_u}^2|}{{\sigma_u}^2} 
    \label{IST(wspd)}
\end{equation}
\begin{equation}
    \mathrm{IST_{wdir}(\degree)} = \mathrm{|wdir_{5min}-wdir_{1hr}|}_{max}
    \label{IST(wdir)}
\end{equation}
In Eq.~\ref{IST(wspd)}, ${\sigma_u^i}^2$ is the streamwise velocity variance for each 5-min block within the 1-hour run and ${\sigma_u}^2$ is the variance for the entire hour. In Eq.~\ref{IST(wdir)}, the notation $|...|_{max}$ represents the maximum difference between the wind direction observed at a 5-minute interval and the wind direction observed at a 1-hour interval among the 12 individual blocks. When the value of $\mathrm{IST}$ is small (close to zero), the streamwise velocity time series can be viewed as stationary. As the value of $\mathrm{IST}$ increases, the non-stationarity effects become significant. According to \cite{moncrieff2004}, $\mathrm{IST_{wspd}}<30\%$ should be achieved to ensure stationarity of the $u$ time series and this metric is employed here.

\item{\textbf{Presence of a Constant \boldmath{$\sigma_w$}}}
An index $I_{\sigma_w}=|\sigma_w - \left<\sigma_w\right>|/\left<\sigma_w\right>$ is defined to measure the deviation of the vertical velocity standard deviation at each level from its depth-averaged value ($\left<\sigma_w\right>$) across the ISL. A small value suggests that the vertical velocity standard deviation is vertically uniform consistent with the AEM. The necessary conditions for the vertical velocity variance to be invariant with $z$ in the ISL may be derived from the mean vertical velocity equation for stationary, planar homogeneous flow in the absence of subsidence at very high Reynolds number. For these idealized conditions, the vertical velocity variance is given by 
\begin{align}
\label{eq:hydropressure}
\frac{\partial \sigma_w^2}{\partial z}=-\left(\frac{1}{\rho}\right)\left(\frac{\partial \overline{P}}{\partial z}\right)-g, 
\end{align}
where $\overline{P}$ is the mean pressure. When $\overline{P}=-\rho g z$ (i.e. hydrostatic), $\partial\sigma_w^2/\partial z=0$ or $\sigma_w^2$ is constant with respect to $z$ within the ISL. That is, the AEM requires $\overline{P}$ to be hydrostatic above and beyond the zero-pressure gradient condition needed to ensure a constant turbulent stress with $z$. The effects of adverse pressure gradients on $C_1$ have already been reported in wind tunnels and pipes \citep{turan1987experimental}.  In the absence of mean pressure gradients, values of $C_1$ varying from 0.90 to 0.92 have been reported. Interestingly, for large adverse pressure gradients, a -1 power law was still reported but values of $C_1$ as high as 17 were computed \citep{turan1987experimental}. These laboratory experiments underscore connections between the aforementioned constant $\sigma_w^2$ with $z$, finite mean pressure gradients, and the numerical values of $C_1$ and $A_1$. 

 \item{\textbf{Reynolds Number}}
As earlier noted, sufficiently large Reynolds numbers are necessary when applying the AEM allowing for self-similarity of turbulent structures \citep{Townsend1976}. Numerous studies highlight the importance of high Reynolds numbers in validating the AEM and observing the expected turbulent behaviour \citep{Marusic2010, Smits2011, Banerjee2013,Huang2022}. Therefore, the dependency of fitted $A_1$ on $Re_{\tau}$ is also investigated. To calculate a bulk Reynolds number analogous to what is reported in laboratory studies, $\delta$ must be estimated. No direct measurement of $\delta$ was available during the experiment and the boundary layer height was estimated from $\delta = C_a u_*/f_c$, where $C_a$ is a dimensionless empirical constant typically varying between 0.1 and 0.3. Here $C_a=0.1$ is used after examining the streamwise energy spectrum as discussed later. The $f_c$ is the Coriolis parameter and is given by $f_c=2\Omega \sin\left(\phi\right)$, where $\Omega$ is the angular velocity of the Earth and $\phi$ is the latitude of the location \citep{zilitinkevich1972}.
\end{enumerate}

\begin{figure}
\centerline{\includegraphics[width=\textwidth]{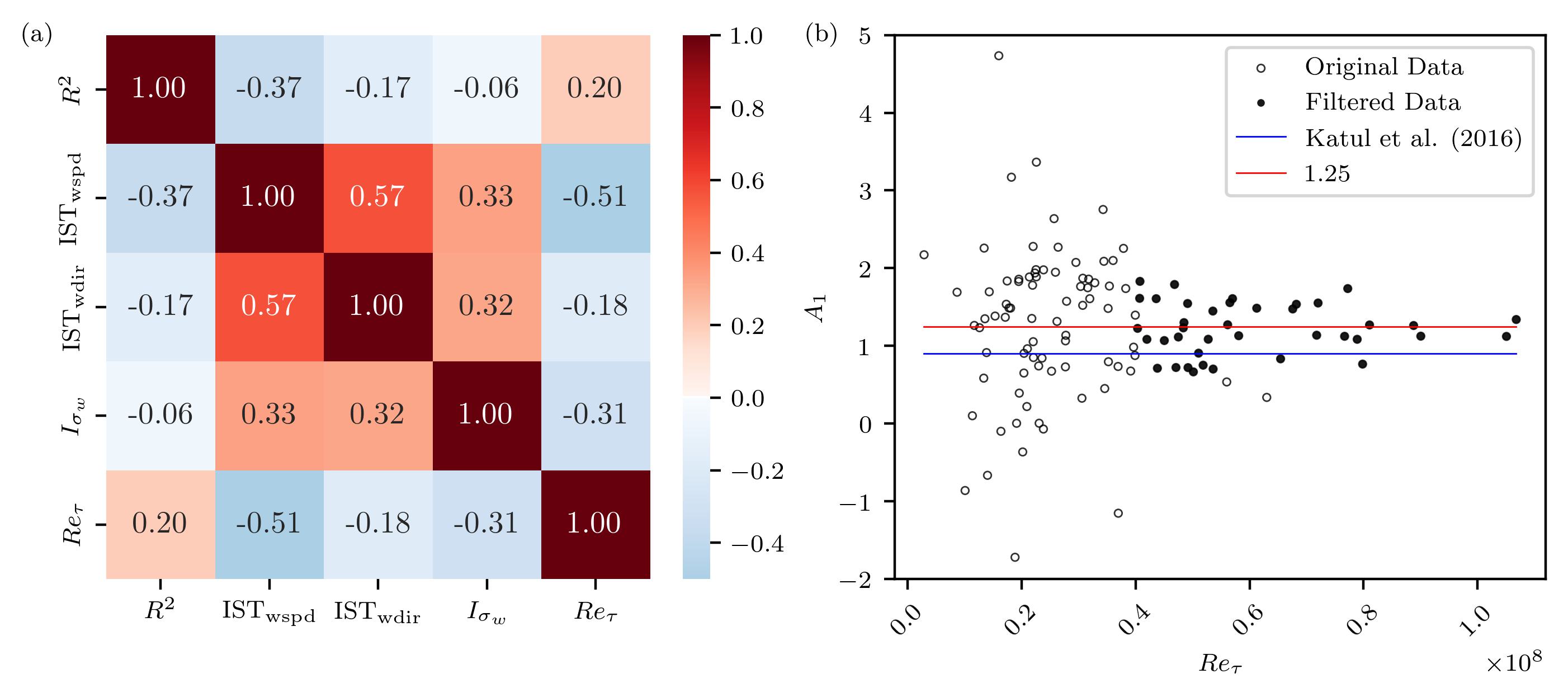}}
\caption{(a) Correlation heatmap between $R^2$ (coefficient of determination from the logarithmic fitting of the streamwise velocity variance), $\mathrm{IST_{wspd}}$ (non-stationarity index of wind speed), $\mathrm{IST_{wdir}}$ (non-stationarity index of wind direction), $I_{\sigma_w}$ (absolute deviation of vertical velocity standard deviation from its mean across the ISL), $Re_{\tau}$ (the bulk Reynolds number). Note that $\mathrm{IST_{wspd}}$, $\mathrm{IST_{wdir}}$ and $I_{\sigma_w}$ are calculated at every height and then averaged across the ISL. (b) Relation between $A_1$ and $Re_{\tau}$. \textit{Open circles} represent the original 120 neutral cases, while \textit{closed circles} denote the 39 cases with $Re_{\tau}>4 \times 10^7$ and $R^2>0.6$. \textit{Blue line} represents the relation between $A_1$ and $Re_{\tau}$ as described in \cite{katul2016}, assuming \( A_1 = C_1 = C_{o,u} \kappa^{-2/3} \), where \( C_{o,u}\) is the Kolmogorov constant for the one-dimensional streamwise velocity spectrum in the inertial subrange (=0.49). \textit{Red line} denotes $A_1=1.25$.}
\label{fig:heatmap_A1_vs_Re}
\end{figure}

Before examining how fitted $A_1$ is affected by these individual factors mentioned above, it is noted that these factors are not entirely independent of each other. Figure~\ref{fig:heatmap_A1_vs_Re}a reveals a moderate negative correlation between \(R^2\) and \(\mathrm{IST_{wspd}}\), indicating that increased non-stationarity in wind speed reduces the quality of the logarithmic fit. A weak negative correlation between \(R^2\) and \(\mathrm{IST_{wdir}}\) suggests a discernible decrease in fit quality with higher wind direction non-stationarity. More dynamically interesting is that higher computed Reynolds numbers ($Re_{\tau}$) are positively correlated with a better fitting quality (\(R^2\)), agreeing with the basic assumption of AEM. Non-stationarity in wind speed (\(\mathrm{IST_{wspd}}\)) is negatively correlated with $Re_{\tau}$, and greater deviations in the vertical velocity standard deviation ($I_{\sigma_w}$) and higher non-stationarity indices tend to occur at lower Reynolds numbers. Overall, a smaller $Re_{\tau}$ reflects, to some extent, the non-stationarity effect and the absence of a constant $\sigma_w$, all of which implicitly affect the fitting quality for the constants of the AEM at very high $Re_{\tau}$. Therefore, the analysis primarily focuses on the role of \(Re_{\tau}\) in describing the variations in fitted $A_1$ values. 

Figure~\ref{fig:heatmap_A1_vs_Re}b illustrates that the derived \(A_1\) values vary appreciably at lower Reynolds numbers but tend to asymptotically converge to a relatively constant range around 1.25 as \({Re}_{\tau}\) increases. This suggests that despite the ASL's Reynolds numbers being on the order of \(10^7\), the fitted \(A_1\) values remain dependent on the Reynolds number. It is important to note that this dependence is not associated with scale separation due to the energy cascade as suggested by previous studies \citep{Marusic2010, Smits2011, Banerjee2013,Huang2022}, but rather results from the combined effects of non-stationarity, the presence or absence of the overlap region, and systematic bias.

When $Re_{\tau}>4 \times 10^7$ is used as the criterion, 41 cases remain, all of which satisfy \(\mathrm{IST_{wspd}}<30\%\). Among these 41 cases, 39 cases have \(R^2>0.6\), with the 2 cases of \(R^2<0.6\) shown as open circles. Therefore, the subsequent analysis focuses on the 39 cases where \(R^2>0.6\) and $Re_{\tau}>4 \times 10^7$.

\begin{table}
\centering
\begin{ruledtabular}
\begin{tabular}{lcccc}
& \multicolumn{2}{c}{\textbf{Before}} & \multicolumn{2}{c}{\textbf{After}} \\ 
    \cmidrule(r){2-3} \cmidrule(r){4-5} 
Variable & Mean & Std & Mean & Std \\
\midrule
$R^2$ & $0.79$ & $0.25$ & $0.88$ & $0.08$ \\
$\mathrm{IST_{wspd}}$ & $0.13$ & $0.11$ & $0.07$ & $0.04$ \\
$\mathrm{IST_{wdir}}$ & $8.03$ & $5.74$ & $7.21$ & $3.48$ \\
$I_{\sigma_w}$ & $0.03$ & $0.02$ & $0.02$ & $0.01$ \\
$Re_{\tau}$ & $3.66 \times 10^7$ & $2.12 \times 10^7$ & $6.06 \times 10^7$ & $1.76 \times 10^7$ \\
\end{tabular}
\end{ruledtabular}
\caption{Comparison of Mean and Std (standard deviation) before and after applying the screening of $Re_{\tau}>4 \times 10^7$ and $R^2>0.6$. Variables are defined as in Figure~\ref{fig:heatmap_A1_vs_Re}.}
\label{table:comparison}
\end{table}

Table~\ref{tab:kappa_A1} and ~\ref{table:comparison} present the results before and after applying the criteria of $Re_{\tau}>4 \times 10^7$ and $R^2>0.6$. After these quality controls are applied, the mean value of $A_1$ changes from 1.25 to 1.22. The mean value of $\kappa$ remains consistent (from 0.37 to 0.38), though with a reduced standard deviation. The $R^2$ values show marked improvement, and the non-stationarity indices decrease substantially. The vertical variability of $\sigma_w$ also decreases from 0.03 to 0.02. 

Post-screening, the patterns of near-neutral occurrence hours and wind directions remain consistent with the previous data. Instances that are infrequent pre-screening become even rarer after the screening, while common instances become more frequent (see Figure~\ref{fig:neutral_hist}c and \ref{fig:neutral_hist}d). The double rotation and planar fit methods show even smaller differences (see Table~\ref{tab:kappa_A1}). The analysis indicates that the variations in $A_1$ are primarily driven by non-stationarity considerations and Reynolds number dependence. However, the variability of $A_1$ (std equal 0.33) remains non-trivial compared to the classical values derived from laboratory experiments despite all the data quality controls imposed on these ASL measurements. This highlights the complexity of ASL flows, necessitating case-by-case investigations, which will be discussed in Section~\ref{sec:subjective}.
\subsubsection{Effect of stability}

\begin{figure}
\centerline{\includegraphics[width=\textwidth]{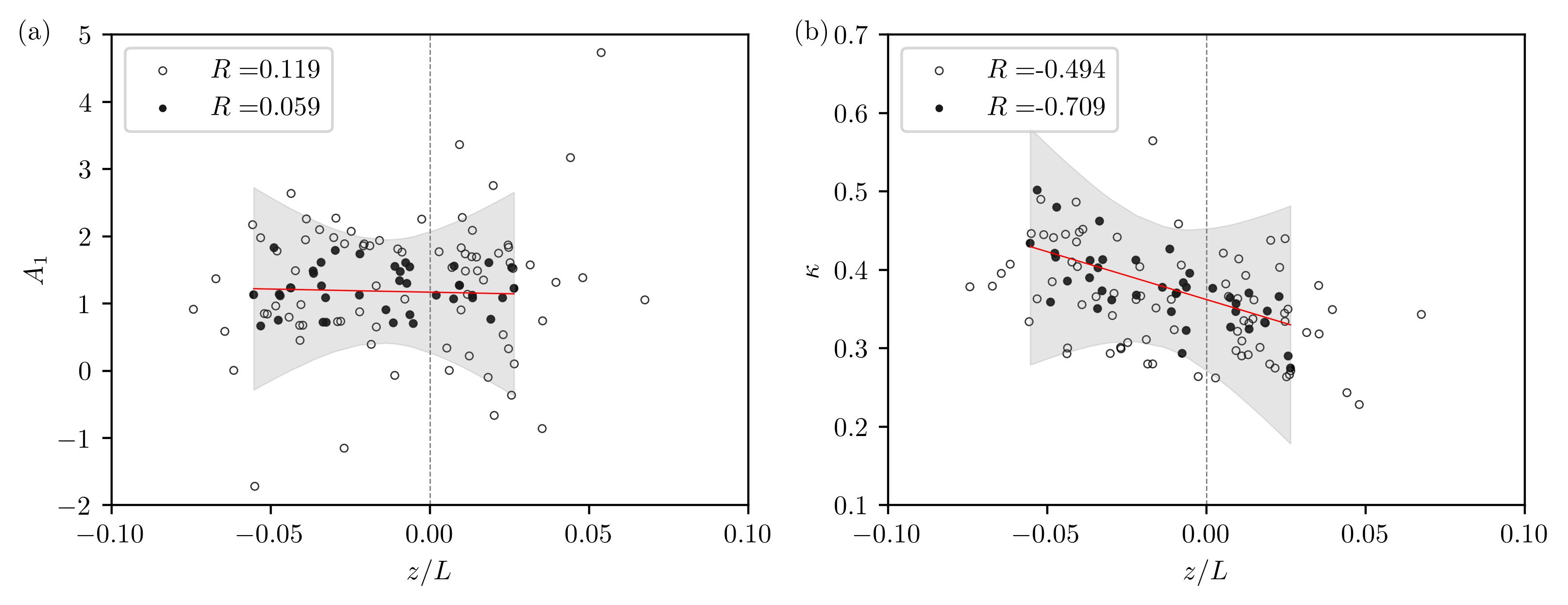}}
\caption{Relation between (a) the fitted $A_1$ and (b) the fitted $\kappa$ and stability parameters $z/L$ of the 120 neutral cases. \textit{Vertical dashed lines} represent $z/L=0$, \textit{solid red lines} denote the linear regression of the data, \textit{grey shadings} denote the 95$^{th}$ confidence level, and \textit{closed circles} denote the 39 cases with $Re_{\tau}>4 \times 10^7$ and $R^2>0.6$. Note that 2 outliers in (a) and 3 outliers in (b) are not shown for better visualization. Insets are the correlation coefficients ($R$).}
\label{fig:kappa_vs_stability}
\end{figure}
In unstable conditions, thermal plumes enhance vertical mixing and modify the turbulent eddy structure. In stable conditions, near-surface turbulence and the residual layer above it progressively decouple with increasing stability, reducing the influence of ABL-scale motions on near-surface turbulence and diminishing the size of streamwise streaks. In those conditions, near-surface turbulence becomes intermittent \citep{dupont2022}. These modifications can influence the streamwise velocity variance and energy spectra, including the $k^{-1}$ scaling \citep{banerjee2015revisiting}. Although $\left|z/L\right|<0.1$ is applied to select near-neutral cases, the selected cases could still be influenced by buoyancy effects (especially the higher $z$ measurements). Discerning these effects is now considered.  

It is observed that sensible heat flux ($|H|$) shows a strong positive correlation with \(Re_{\tau}\) (correlation coefficient of 0.65), indicating that higher sensible heat flux tends to be associated with higher Reynolds numbers. After the data quality control, the sensible heat flux remains high and variable both before and after screening, posing a challenge in achieving both high Reynolds numbers and minimal buoyancy effects. In the ABL, this correlation is not surprising as stronger mean winds usually occur during daytime conditions when surface heating is large.  Figure~\ref{fig:kappa_vs_stability}a illustrates the relation between the stability parameter and the fitted \(A_1\). The screening criteria of \( Re_{\tau}>4 \times 10^7 \) and \( R^2 > 0.6 \) exclude data with stability parameters greater than 0.026 (positive, stable) and less than -0.05 (negative, unstable). The remaining data are roughly evenly distributed across the range (-0.05, 0.025) with no significant trend. This suggests that there is little stability effect on $A_1$ after screening when stability is quantified using the local Obukhov length and the local measurement height. This finding is not sensitive to the initial choice of $\left|z/L\right|<0.1$, which justifies its use to indicate near-neutral conditions.

In contrast, the influence of stability conditions on the fitted values of $\kappa$ is visible in Figure~\ref{fig:kappa_vs_stability}b. Even after applying the stability constraint, the fitted $\kappa$ values remain influenced by variations in atmospheric stability despite the near-neutral filter, consistent with other prior experiments \citep{Andreas2006}. It is possible to correct the stability effect on $\kappa$ by employing MOST and the Businger-Dyer relation for the stability correction function \citep{businger1988note}. After correcting the stability effects, the median value of $\kappa$ estimated from the dataset analyzed here is 0.36 (not shown). This aligns with Figure~\ref{fig:kappa_vs_stability} where the vertical dashed line intersects the regression line roughly at $\kappa = 0.36$. 

\subsection{Streamwise velocity energy spectrum}

 \subsubsection{Ensemble streamwise velocity energy spectrum}
\begin{figure}
\centerline{\includegraphics[width=\textwidth]{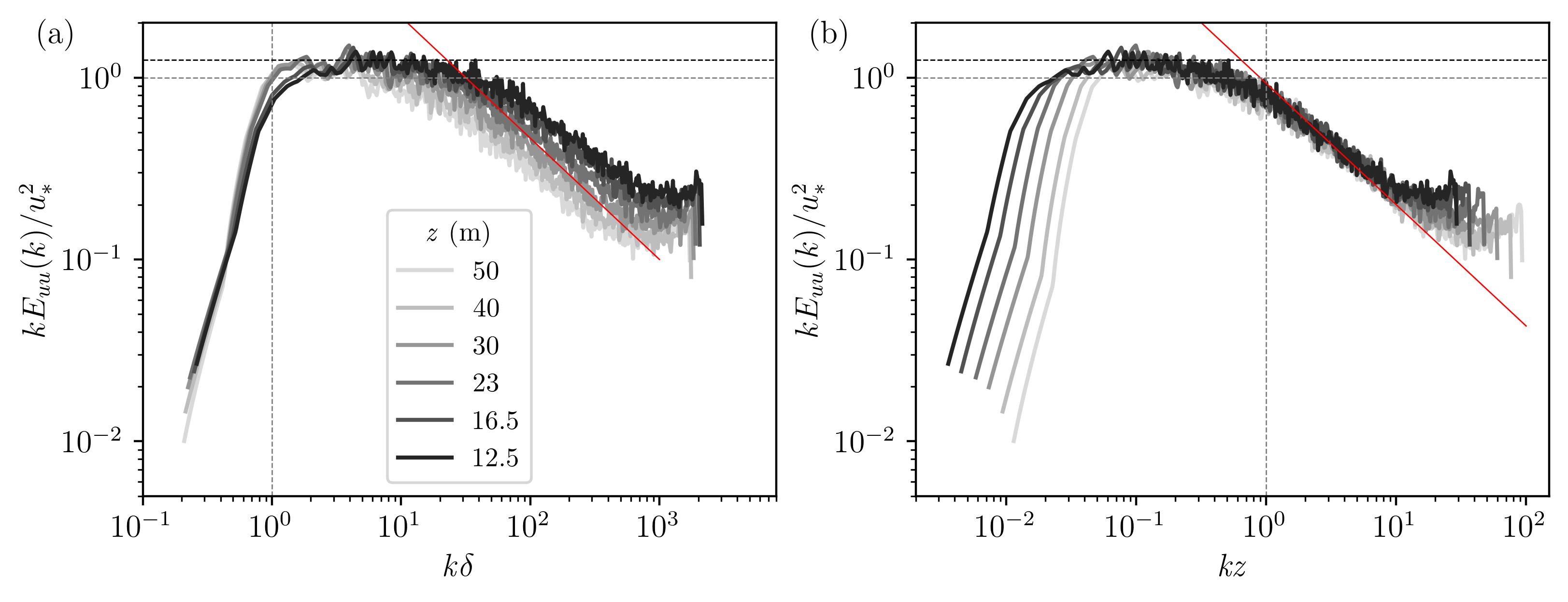}}
\caption{The ensemble average of the pre-multiplied streamwise velocity spectrum over 39 post-screening cases from 12.5 (lightest shade) to 50 metres (darkest shade) with (a) $\delta$-scaling and (b) $z$-scaling. \textit{Horizontal dashed lines} mark 1 and 1.25. \textit{Vertical dashed lines} indicate $k\delta$ or $kz$ equals 1. \textit{Solid red line} denotes the \(-2/3\) scaling.}
\label{fig:ensemble}
\end{figure}
 
The pre-multiplied streamwise velocity energy spectra are computed within the ISL using the Welch method \citep{welch1967} while applying a Hann window with a $2^{14}$ window size and $50\%$ overlap between consecutive windows. The power spectral density is then interpolated onto 500 logarithmically spaced frequencies, spanning from $10^{-4}$ to 5 Hz (Nyquist frequency). Figure~\ref{fig:ensemble} shows the ensemble average (i.e., average over 39 post-screening cases) of the pre-multiplied $u$-spectra. The streamwise wavenumber $k$ is calculated based on Taylor's hypothesis as $k=2\pi f/\overline{u}$, where $f$ is the frequency. Due to the normalization of wavenumbers by $z$, the pre-multiplied spectra at different heights collapse well in the inertial subrange \citep{kaimal1994atmospheric}. The pre-multiplied \(u\)-spectrum rises at the tail due to high-frequency noise. However, this high-frequency noise does not significantly affect the streamwise velocity variance and is neglected. 

A \(k^{-2/3}\) power-law scaling is evident in the inertial subrange wavenumber region, consistent with Kolmogrov's theory. More interestingly, a plateau of about 1 to 1.25 begins roughly at $k\delta = 1$ (see Figure~\ref{fig:ensemble}a), consistent with the onset of the $k^{-1}$ scaling predicted by the AEM, suggesting that the estimation of $\delta = C_a u_*/f_c$ with $C_a=0.1$ is plausible. Furthermore, this plateau occurs within the range $\mathcal{O}(0.1) < kz < \mathcal{O}(1)$ as shown in Figure~\ref{fig:ensemble}b, indicating a robust \(k^{-1}\) power-law scaling for $E_{uu}(k)$. However, the \(k^{-1}\) scaling region is just under a decade in extent prompting the question of whether individual cases show clear \(k^{-1}\) scaling and whether the plateau values ($C_1$) correspond to the fitted $A_1$ values, which will be discussed next.

\subsubsection{Additional quality control}
\label{sec:subjective}

\begin{figure}
\centerline{\includegraphics[width=\textwidth]{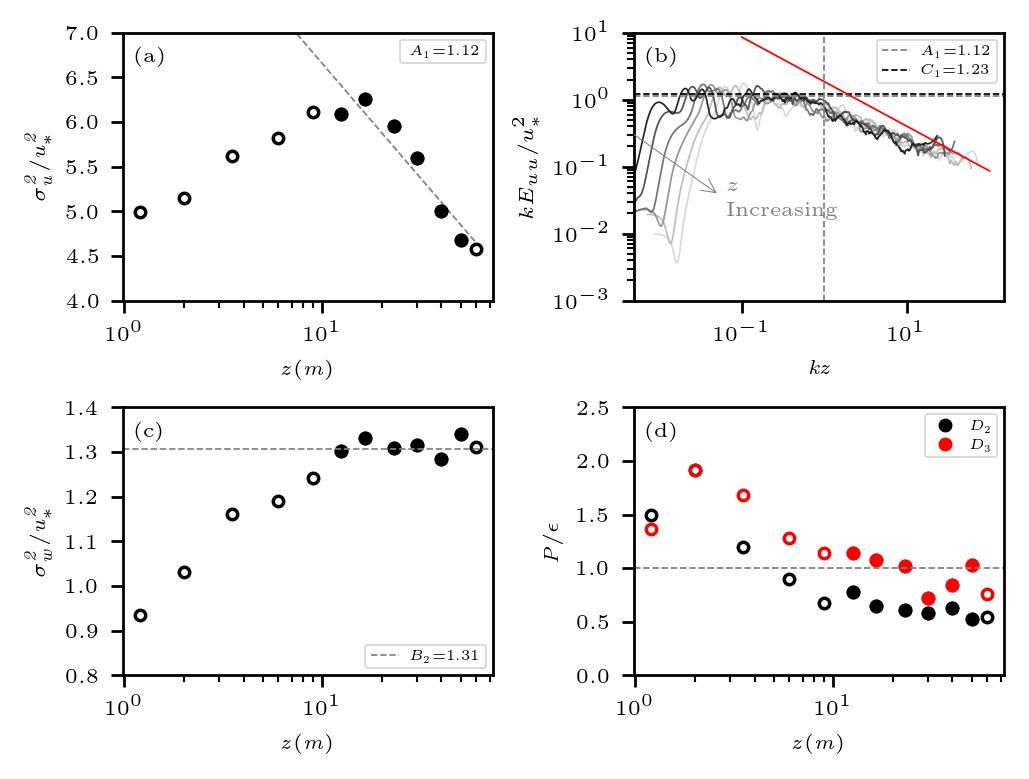}}
\caption{ An example of a benchmark case collected on Mar. 29, 2021, at 0400LT. (a) Profile of $\sigma_u^2$. \textit{Dashed grey line} denotes the linear regression of the data within the ISL. Insets indicate the fitted $A_1=1.12$ of this case. (b) The pre-multiplied streamwise velocity spectrum normalized by $u_*$ ($kE_{uu}/u_*^2$) against wavenumber normalized by height ($kz$). The arrow represents increasing $z$. \textit{Vertical dashed line} marks $kz=1$. \textit{Horizontal dashed lines} mark the fitted $A_1=1.12$ (in grey) and $C_1=1.23$ (in black) estimated by the depth-average of the 95$^{th}$ percentile of $kE_{uu}/u_*^2$. \textit{Solid red line} denotes the \(-2/3\) scaling. 
(c) Profile of $\sigma_w^2$. \textit{Dashed grey line} represents the estimated $B_2=1.31$. (d) The ratio of production rate $p$ to dissipation rate $\epsilon$ against $z$. $p$ is calculated by the third-order polynomial fitting of the mean wind profile, and $\epsilon$ is calculated using both the second-order structure function (in black) and the third-order structure function (in red). \textit{Dashed grey line} represents where $P/\epsilon=1$. \textit{Closed circles} indicate the selected inertial sublayer.}
\label{fig:subjective}
\end{figure}

To analyze the linkage between the logarithmic behaviour of streamwise velocity variance and the \(k^{-1}\) scaling in the streamwise energy spectra (\(E_{uu}\)), each of the 39 remaining cases is examined individually. $C_1$ are approximated by the 95$^{th}$ percentile values of \(kE_{uu}/u_*^2\) for simplicity. Although this is an approximation, it enables us to explore a general correlation between $A_1$ and $C_1$. 

The profile of the vertical velocity variance and the TKE budget are also examined. In the ISL, the TKE budget simplifies to a balance between the mechanical production ($P$) and the viscous dissipation rate ($\epsilon$). The production rate is determined by:
\begin{equation}
    P=-\overline{u'w'}\frac{\partial U}{\partial z}=u_*^2\frac{\partial U}{\partial z},
\end{equation}
where \({\partial U}/{\partial z}\) is approximated in two ways: first, using \({u_*}/{\kappa z}\) based on the log law, where $\kappa$ is the fitted value; second, using a third-order polynomial fitting of the mean wind profile over all 12 levels. These two methods yield similar results (as shown in Figure~\ref{fig:p_E_scatter}a in the appendix). The production rate obtained using the second method is used in the analysis of the TKE budget. The dissipation rate ($\epsilon$) is determined by both the second-order structure functions ($D_2(r)$) and the third-order structure functions ($D_3(r)$), respectively. At very high Reynolds numbers, these are given as follows
\begin{subequations}
    \begin{equation}
    D_2(r)=C_2\epsilon^{\frac{2}{3}}r^{\frac{2}{3}},
    \end{equation}

    \begin{equation}
    D_3(r)=-\frac{4}{5}\epsilon r,
    \end{equation}
\end{subequations}
where \( C_2 \) is a coefficient ($\approx 1.97$ for one-dimensional wavenumber), and $r$ represents the separation distance in the streamwise direction, set to be the minimum between 1 meter and $z/2$~\citep{chamecki2004local}.  Prior studies have shown that the ratio of $D_2(r)$ in the vertical to the streamwise direction most closely matches the isotropic value of 4/3 when $0.63$ m $\leq r \leq$ 1.0 m \citep{chamecki2004local}. A similar eddy size range was also identified in other ASL experiments \citep{katul1997energy}. Figure~\ref{fig:p_E_scatter}b shows that the dissipation rates derived from the second-order structure function are slightly smaller than their third-order counterpart. In the following analysis, both dissipation rates are used to assess the sensitivity of the AEM to TKE dissipation rate estimation.

Three cases (out of 39) exhibit anomalous \(\sigma_u^2\) profiles (Group 1; see Figure~\ref{fig:anomalous sigma_u}), while nine cases show anomalous \(\sigma_w^2\) profiles with clear trends within the ISL (Group 2; see  Figure~\ref{fig:anomalous sigma_w}). Seven cases display anomalous \(E_{uu}\) profiles, where the pre-multiplied streamwise velocity spectrum presents two or more peaks in the production range (Group 3; see  Figure~\ref{fig:anomalous Euu}), suggesting the contribution of eddies other than the attached eddies. Five cases have an anomalous TKE balance (Group 4; see  Figure~\ref{fig:anomalous TKE}), with the ratio of production to dissipation rates showing clear trends in the ISL, likely indicating non-equilibrium turbulence, potentially influenced by sudden changes in wind speed, temperature gradients, or other factors. These anomalous behaviours of turbulent properties can coincide, though this overlap is not fully explored in the categorization. These anomalies may be due to local topographic features, instrumentation issues, or transient meteorological phenomena. It is acknowledged that the categorization is subjective to the authors' choices. Therefore, this section serves only to provide a qualitative conclusion.  

Ultimately, nine cases are flagged as high-quality data (Group 5), serving as benchmarks for understanding the connection between streamwise velocity variance and the streamwise velocity energy spectrum. Figure~\ref{fig:subjective} provides an example of this high-performance group on Mar. 29, 2021, at 0400LT. The profile of $\sigma_u^2$ exhibits a logarithmic behaviour, with a fitted $A_1$=1.12 (Figure~\ref{fig:subjective}a). Similar to the ensemble average, the pre-multiplied $E_{uu}$ reach a plateau in the range of $\mathcal{O}(0.02)<kz<\mathcal{O}(1)$. The estimated $C_1$ is 1.23, marginally higher than the fitted $A_1$. At the inertial subrange, the streamwise energy spectra align well with a $k^{-2/3}$ scaling (Figure~\ref{fig:subjective}b) as expected. The vertical velocity variance remains constant with $z$ in the ISL (Figure~\ref{fig:subjective}c), with the $\sigma_w^2/u_*^2$ values centred around 1.3. The ratio $P/\epsilon$ derived from both $D_2(r)$ and $D_3(r)$ is invariant with height within the ISL. However, $P/\epsilon$ based on $D_2(r)$ is smaller than that based on $D_3(r)$, with the latter being near-unity.


\begin{figure}
    \centering
  \includegraphics[width=\linewidth]{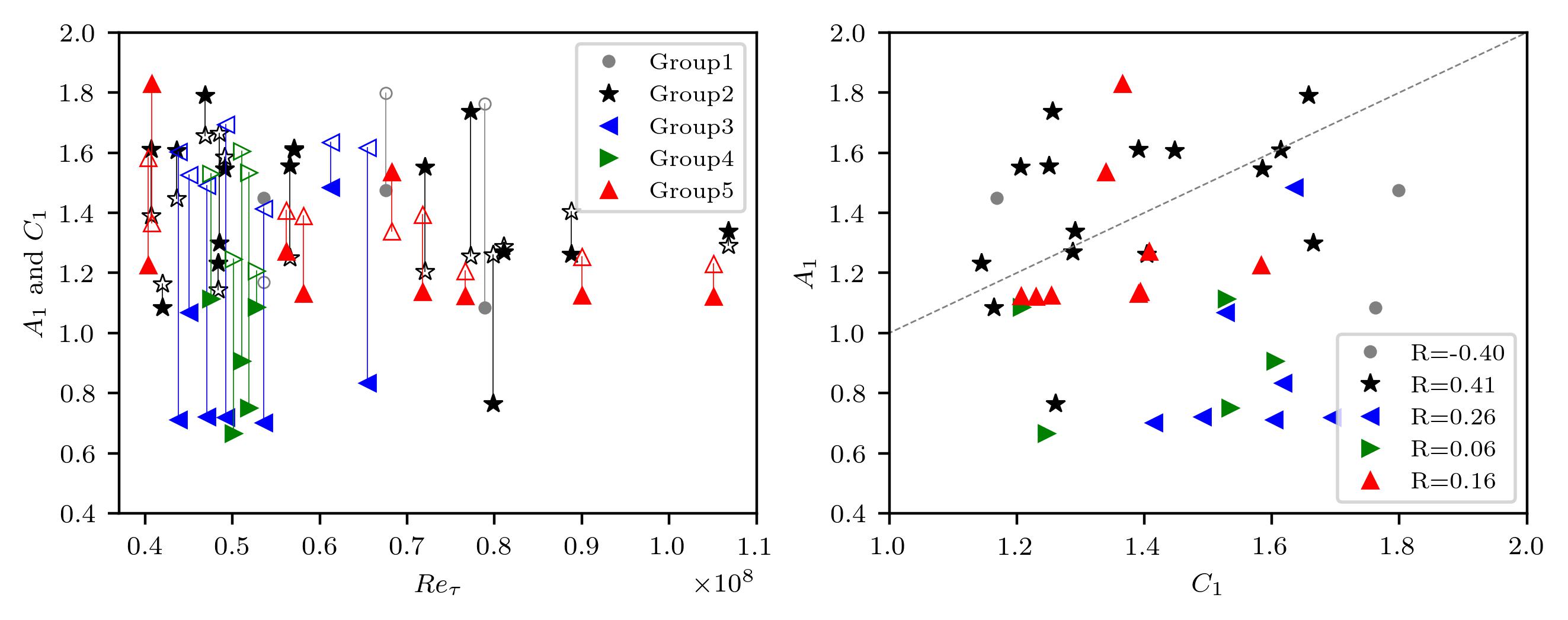}
    \caption{(a) Relation between the fitted $A_1$ (closed symbols) as well as the estimated $C_1$ (open symbols) against $Re_{\tau}$. Group 1: 3 cases with anomalous $\sigma_u^2$ profiles denoted by \textit{grey circles}. Group 2: 9 cases with anomalous $\sigma_w^2$ profiles denoted by \textit{black stars}. Group 3: 7 cases with anomalous $E_{uu}$ profiles denoted by \textit{blue left-pointing triangles}. Group 4: 5 cases with anomalous TKE budget balance denoted by \textit{green right-pointing triangles}. Group 5: 9 cases with high performance denoted by \textit{red upward triangles}. (b) Relation between the estimated $C_1$ and the fitted $A_1$. \textit{Grey dashed line} denotes the 45\degree slope. Insets are the correlation coefficients (R) for each group.}
    \label{fig:95th_vs_A1}
\end{figure}

To investigate whether a $C_1=A_1$ relation can be observed in the adiabatic ASL, the estimated $A_1$ and $C_1$ of the 39 post-filtered cases were grouped based on these additional quality controls and plotted against $Re_{\tau}$ (see Figure~\ref{fig:95th_vs_A1}a). Each vertical line connects the open and closed symbols for the same \( Re_{\tau} \), indicating the difference between the corresponding \( A_1 \) and \( C_1 \) for each case within the groups.
In Groups 1, 3, and 4, the estimated $C_1$ values differ substantially from their corresponding $A_1$ values. Groups 2 and 5 exhibit a spread of $A_1$ and $C_1$ values across different $Re_{\tau}$. Interestingly, while there is some variation in these values at lower $Re_{\tau}$, they appear to converge at higher $Re_{\tau}$, which seems to suggest that both $A_1$ and $C_1$ may depend on $Re_\tau$ and other atmospheric factors rather than being constants within the ASL. Figure~\ref{fig:95th_vs_A1}b further explore the relation between $A_1$ and $C_1$, showing a weak-to-no correlation between the fitted $A_1$ and the estimated $C_1$ values among all the groups. Therefore, the data in this work do not support the $C_1=A_1$ relation proposed by AEM.
 
\section{Conclusions}
The logarithmic behaviour of the streamwise velocity variance ($\sigma_u^2$) in the near-neutral ASL predicted by the AEM is explored using eddy-covariance observations from a 62-m tower located in the Eastern Snake River Plain, Idaho, US. Additionally, the study examines the relation between the logarithmic behaviour of the $\sigma_u^2$ and the $k^{-1}$ scaling in the streamwise energy spectra.

The analysis indicates that the streamwise velocity variance in the near-neutral ASL follows a logarithmic profile within the inertial sublayer. The fitted value of the Townsend-perry coefficient ($A_1$) is sensitive to weak non-stationarity. To bypass this issue, some filtering is needed to remove any non-linear trends. In contrast, the fitted von K\'arm\'an constant ($\kappa$) is not as sensitive to these effects but is influenced by atmospheric stability conditions. 

The variability in the fitted value of $A_1$ is largely attributed to non-stationarity,  Reynolds number dependence, and quality of the fitting process. After controlling for these factors, 39 hours of data remain and their fitted \(A_1\) values converge to the range of 1 to 1.25. This variability remains non-trivial (mean = 1.22, std = 0.33, min = 0.67, max = 1.83).  

A further selection of 9 cases with canonical \(\sigma_u^2\) and \(\sigma_w^2\) profiles, \(E_{uu}\) shapes, and expected TKE budget balance reveals that, in these cases, the pre-multiplied \(E_{uu}\) reaches a plateau in the production range, spanning roughly a decade, equivalent to a $k^{-1}$ scaling in \(E_{uu}\). Both the fitted $A_1$ and the estimated $C_1$ display some scatter at lower $Re_{\tau}$, but as $Re_{\tau}$ increases, a relatively tighter and more consistent relation between $A_1$ and $C_1$ emerges. However, no direct evidence supports the $C_1=A_1$ relation in these cases. 

These findings suggest that ASL turbulence does obey the AEM with coefficients commensurate with those reported for very high Reynolds number laboratory experiment. It also highlights the complexity of ASL, where non-stationary effects, large-scale and very large-scale eddies, and local topographic features complicate the identification of the inertial sublayer in streamwise velocity variance profiles and the $k^{-1}$ scaling range in the streamwise energy spectrum. Future studies seek to explore two inter-related aspects of ISL flow statistics: the m-higher order moments of $(\sigma_u/u_*)^m$ for near neutral conditions and the role of atmospheric stability on the scaling laws of $\sigma_u$ and its spectral exponents at large scales.
 
\begin{acknowledgements}
Y. Qin and D. Li acknowledge support from the U.S. National Science Foundation (NSF-AGS-1853354). 
G. Katul acknowledges support from the U.S. National Science Foundation (NSF-AGS-2028633) and the Department of Energy (DE-SC0022072). H. Liu acknowledges support from the U.S. National Science Foundation (NSF-AGS-1853050).
We thank Zhongming Gao at Sun Yat-sen University for his assistance with data collection and preprocessing.
\end{acknowledgements}

\appendix

\section{\label{appA}Supplementary figures}
    \setcounter{figure}{0}
    \renewcommand{\thefigure}{A\arabic{figure}}

\begin{figure}   \centerline{\includegraphics[width=\textwidth]{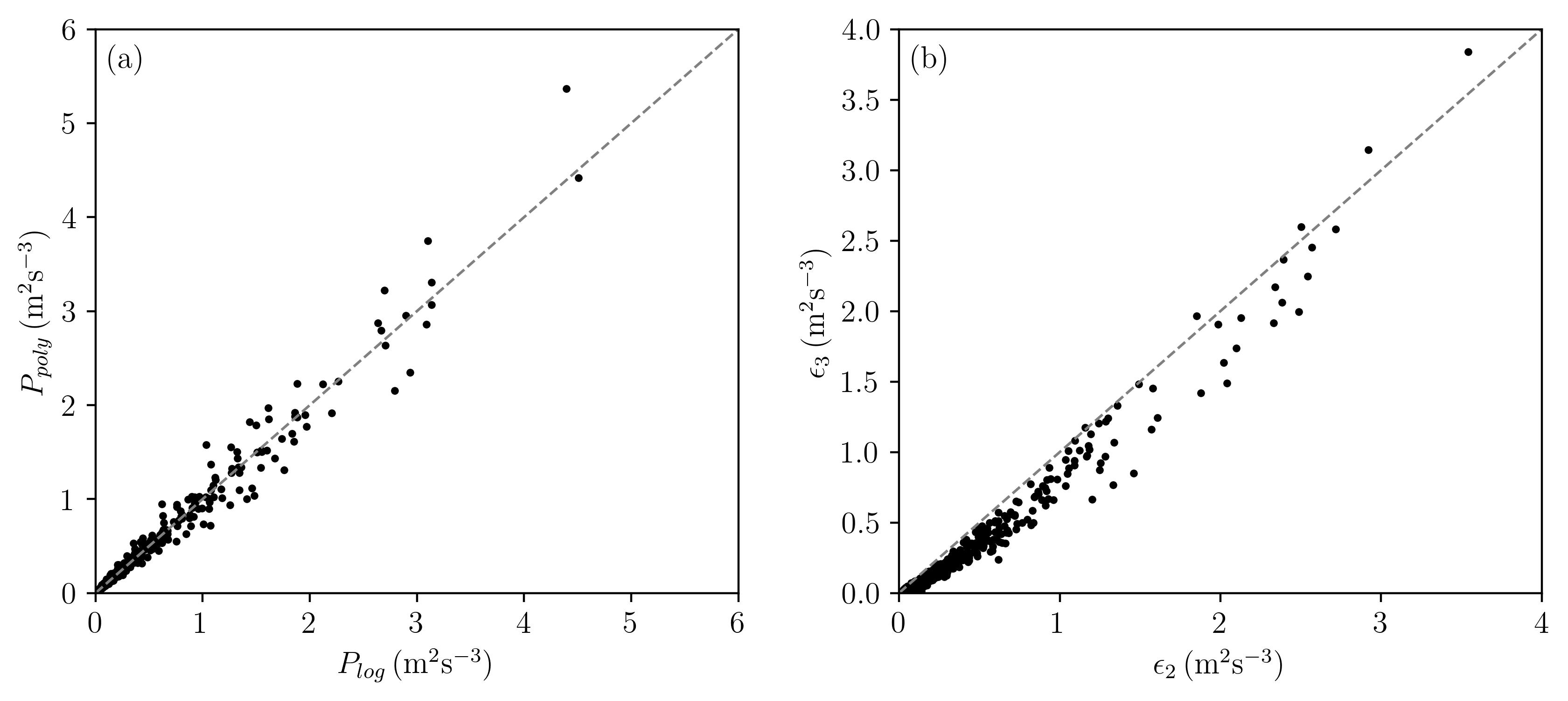}}
    \caption{(a) Comparison between the mechanical production rate calculated using the log law ($P_{log}$) and the polynomial fitting to mean velocity data ($P_{poly}$). (b) Comparison between the turbulent kinetic energy dissipation rate calculated using the second-order structure function ($\epsilon_2$) and the third-order structure function ($\epsilon_3$). \textit{Grey dashed lines} represent perfect agreement.}
    \label{fig:p_E_scatter}
\end{figure}

\begin{figure}
\centerline{\includegraphics[width=\textwidth]{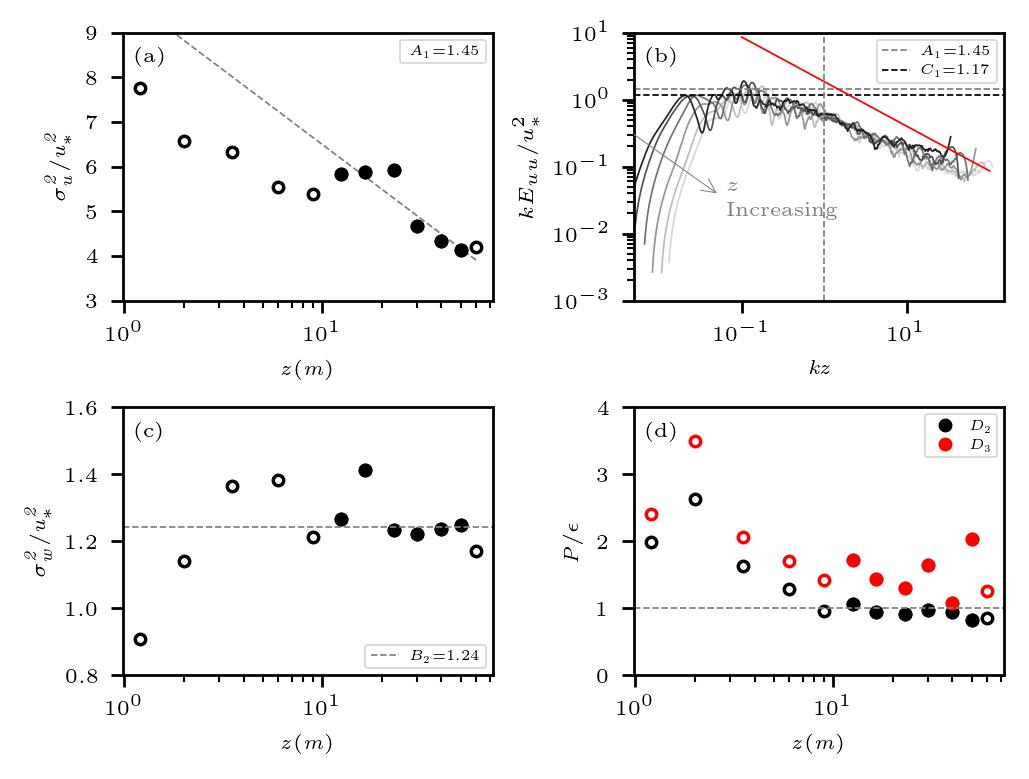}}
\caption{Example of anomalous $\sigma_u^2$: similar to Figure~\ref{fig:subjective} with data from Mar. 6, 2021, at 1500LT.}
\label{fig:anomalous sigma_u}
\end{figure}

\begin{figure}
\centerline{\includegraphics[width=\textwidth]{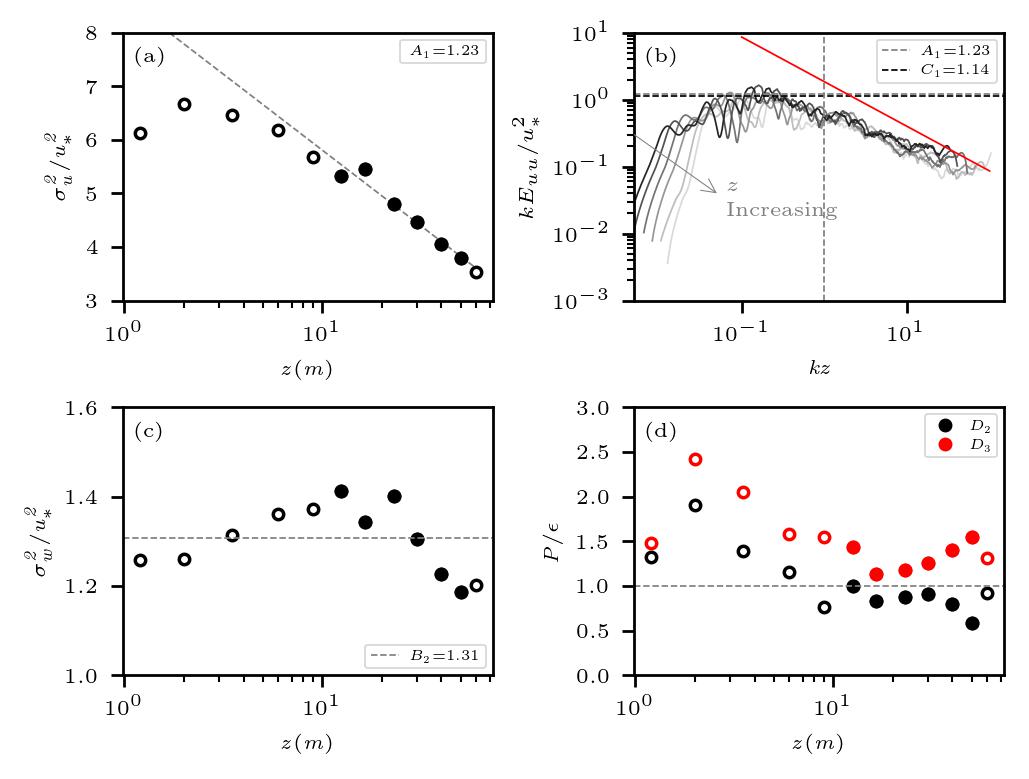}}
\caption{Example of anomalous $\sigma_w^2$: similar to Figure~\ref{fig:subjective} with data from Apr. 14, 2021, at 1100LT.}
\label{fig:anomalous sigma_w}
\end{figure}

\begin{figure}
\centerline{\includegraphics[width=\textwidth]{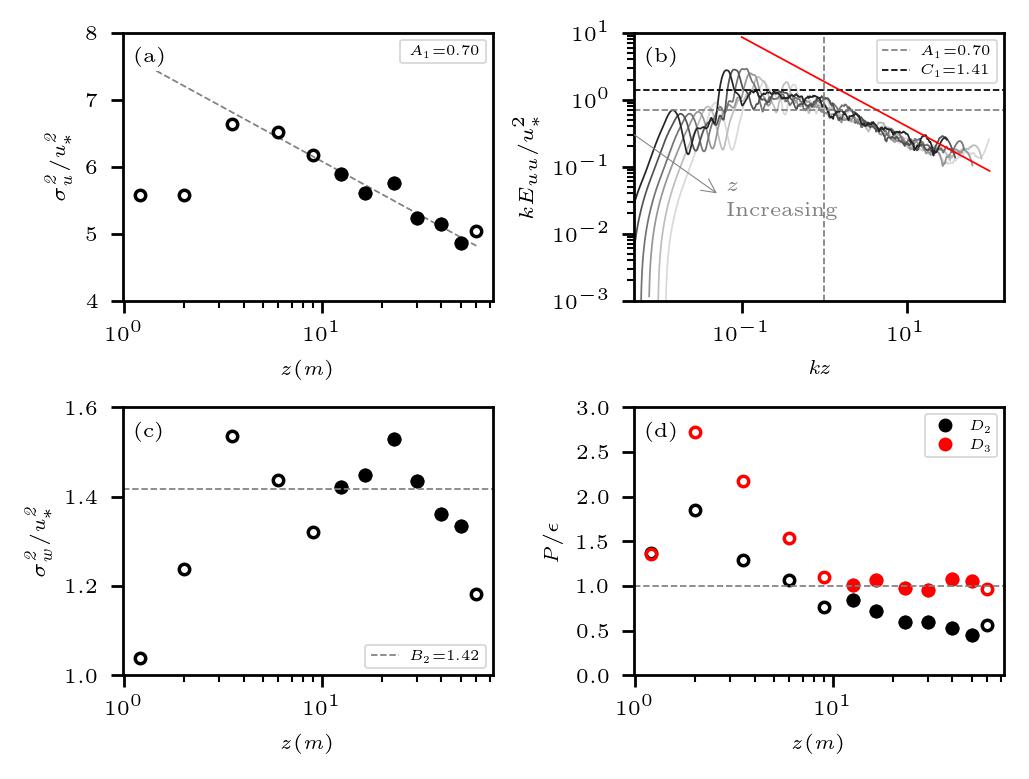}}
\caption{Example of anomalous $E_{uu}$: similar to Figure~\ref{fig:subjective} with data from Oct. 10, 2020, at 1600LT.}
\label{fig:anomalous Euu}
\end{figure}

\begin{figure}
\centerline{\includegraphics[width=\textwidth]{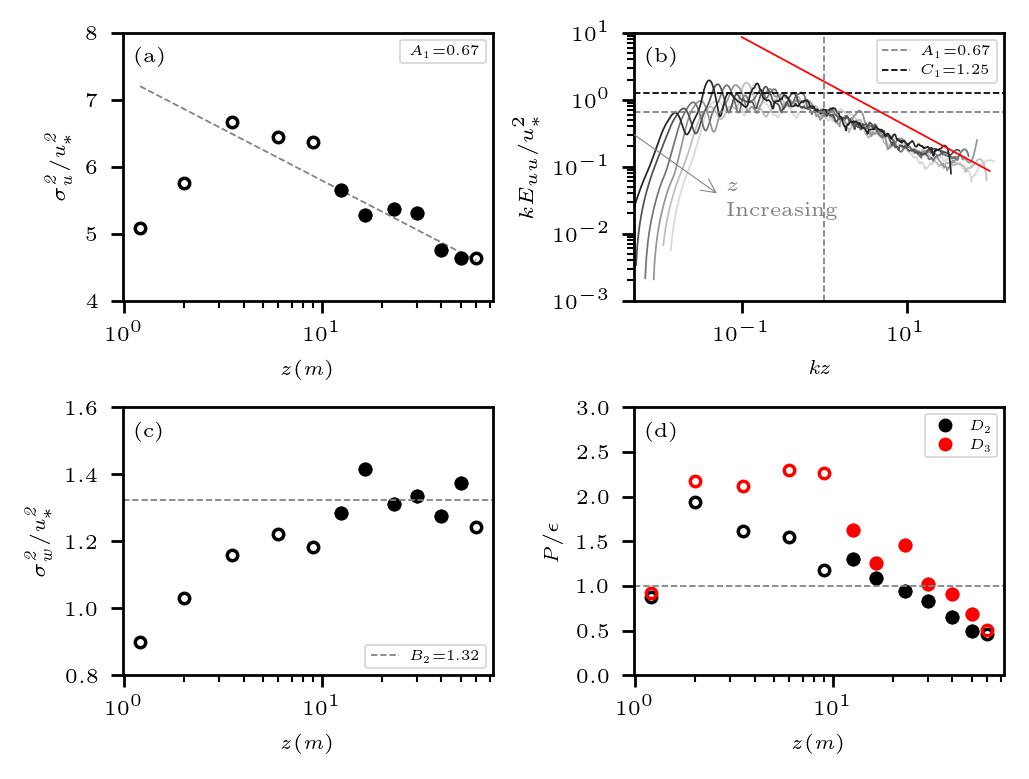}}
\caption{Example of anomalous TKE budget balance: similar to Figure~\ref{fig:subjective} with data from Nov. 19, 2020, at 1400LT.}
\label{fig:anomalous TKE}
\end{figure}

\bibliography{apssamp}

\end{document}